\documentclass[11pt,aps,prd,nofootinbib]{revtex4}
\usepackage{graphicx}
\usepackage{dcolumn}
\usepackage{bm}
\usepackage{epsfig}
\usepackage{dcolumn}
\usepackage{morefloats}
\def\be{\begin{equation}}
\def\ee{\end{equation}}
\def\bea{\begin{eqnarray}}
\def\eea{\end{eqnarray}}\def\nn{\nonumber}
\def\gsim{\ \rlap{\raise 2pt\hbox{$>$}}{\lower 2pt \hbox{$\sim$}}\ }
\def\lsim{\ \rlap{\raise 2pt\hbox{$<$}}{\lower 2pt \hbox{$\sim$}}\ }
\def\dslash{\kern-4pt \not{\hbox{\kern-2pt $\partial$}}}
\def\pslash{\not{\hbox{\kern-2pt p}}}

\def\pmue{{{P_{\mu e}} }}
\def\pmumu{{{P_{\mu \mu}} }}

\newcommand{\dcp}{\delta_{CP}}
\newcommand{\nova}{NO$\nu$A\ }

\newcommand{\dmmm}{\Delta_{\mu\mu} }
\newcommand{\tmm}{\theta_{\mu\mu}}

\begin{document}
\DeclareGraphicsExtensions{.eps,.ps}


\title{ Evidence for leptonic CP phase from \nova, T2K and ICAL: A chronological progression}



\author{Monojit Ghosh}
\email[Email Address: ]{monojit@prl.res.in}
\affiliation{
Physical Research Laboratory, Navrangpura,
Ahmedabad 380 009, India}
 
\author{Pomita Ghoshal}
\email[Email Address: ]{pomita@prl.res.in}
\affiliation{
Physical Research Laboratory, Navrangpura,
Ahmedabad 380 009, India}
 
\author{Srubabati Goswami}
\email[Email Address: ]{sruba@prl.res.in}
\affiliation{
Physical Research Laboratory, Navrangpura,
Ahmedabad 380 009, India}

\author{Sushant K. Raut}
\email[Email Address: ]{sushant@prl.res.in}
\affiliation{
Physical Research Laboratory, Navrangpura,
Ahmedabad 380 009, India}

\begin{abstract}

We study the synergy between the long-baseline (LBL) experiments
\nova and T2K and
the atmospheric neutrino experiment ICAL@INO for obtaining the first hint of 
CP violation in the lepton sector. We also 
discuss how precisely the  
leptonic CP phase ($\dcp$) can be measured by these experiments.   
The CP sensitivity is first described at the level of oscillation
probabilities, discussing its dependence on the parameters -- 
$\theta_{13}$, mass hierarchy and $\theta_{23}$. In particular, 
we discuss how the precise knowledge or lack thereof of these parameters can affect the CP sensitivity of LBL experiments. 
We follow a staged approach and analyze the $\dcp$ sensitivity that 
can be achieved at different points of time over the next 15 years 
from these LBL 
experiments alone and/or in conjunction with ICAL@INO.  
We find that the CP sensitivity of \nova/T2K is 
enhanced due to the 
synergies between the  different channels and between the two experiments. 
On the other hand the lack of knowledge of hierarchy and octant makes the 
CP sensitivity poorer for some parameter ranges. 
Addition of ICAL data to T2K and \nova  can exclude these spurious 
wrong-hierarchy and/or wrong-octant solutions and 
cause a significant increase in the range of $\dcp$ values for which 
a hint of CP  violation can be achieved. In fact in parameter regions 
unfavourable for \nova/T2K, we may get the first evidence of CP violation 
by adding  the ICAL data  to these. Similarly the precision with which $\dcp$ 
can be measured also improves with inclusion of ICAL data. 

\end{abstract}
\maketitle

\section{Introduction}

In the present status of neutrino oscillation physics, a fair amount of
knowledge about the oscillation parameters has been gained from
solar, atmospheric, accelerator and reactor experiments.
In the standard 3-flavour scenario there are 6 parameters governing the
oscillation of the neutrinos. These are the three mixing angles ---
$\theta_{12}$, $\theta_{23}$, $\theta_{13}$,
two mass squared differences  --
$\Delta m^2_{31},~\Delta m^2_{21}$ ($\Delta m^2_{ij} = m_i^2 - m_j^2$)
and the Dirac CP phase $\dcp$.  
Among these the unknown parameters are: 
(i) the  
sign of $\Delta m^2_{31}$ 
($\Delta m^2_{31} >0$  corresponds to Normal Hierarchy (NH);   
$\Delta m^2_{31}<0$ corresponds to Inverted Hierarchy (IH) ) 
(ii) the octant of $\theta_{23}$ ($\theta_{23} > 45^\circ$  
corresponding to Higher Octant (HO) 
or $\theta_{23} < 45^\circ$ corresponding to Lower Octant (LO)).   
(iii) the CP phase $\dcp$; a value of this parameter different from 0
or $180^\circ$ would signal CP violation in the lepton sector.  
\footnote{The current best-fit value of $\theta_{23}$ is in the 
lower octant but at 2$\sigma$ the higher octant values 
are allowed. Also reference \cite{lisi-global} shows a preference 
 for $\dcp =180^\circ$ though this is not statistically significant.
Recently a 1$\sigma$ hint towards NH has been reported in 
\cite{Zhang:2013rta}.}

CP violation has been observed in the quark sector and this can be explained 
by the complex phase of the CKM matrix 
\cite{Christenson:1964fg,Aubert:2001nu,Abe:2001xe}. 
The origin of this could be complex Yukawa couplings and/or complex 
vacuum expectation values of the Higgs field \cite{Lee:1973iz,Lee:1974jb}. 
In such cases it is plausible that there can be a complex phase analogous to 
the CKM phase,  
in the leptonic mixing matrix as well
\footnote{In the quark sector there is only one CP phase associated with the 
CKM matrix. However if neutrino are Majorana particle then the 
leptonic mixing matrix contains three phases in general.  
However oscillation experiments are sensitive only to the 
Dirac CP phase $\dcp$.}. 
This can lead to 
CP violation in the lepton sector \cite{leptoncpv}.  
However experimental detection of this phase is necessary to establish 
this expectation on a firm footing. 
The determination of the leptonic CP phase 
is interesting not only in the context of fully
determining the MNSP mixing matrix but also because it could be responsible
for the observed matter-antimatter asymmetry  through the mechanism of 
leptogenesis \cite{Joshipura:2001ui, Endoh:2002wm}. 

Since $\dcp$ occurs with the mixing angle $\theta_{13}$ in the MNSP matrix, the
recent measurement of a non-zero and moderately large value of 
this angle by reactor  
and accelerator experiments is expected to be conducive for the measurement of 
$\dcp$.  
The current best-fit value of $\theta_{13}$ from global oscillation analyses
is  
$\sin^2 2\theta_{13} \approx 0.10 \pm 0.01$~
\cite{lisi-global,valle-global,msg_global}.
If $\theta_{13}$ was very small then any measurement of $\dcp$ would have 
required high intensity sources \cite{iss}. 
However the moderately large value of 
$\theta_{13}$ makes it  worthwhile to explore whether $\dcp$ 
can be measured and  any evidence 
of CP violation can be obtained by the current and upcoming experiments using 
conventional beams.
Many recent studies have investigated this issue
\cite{minakata2,uma_largexp,nova_reopt2,novat2k,coloma} 
in the context of the  LBL experiment T2K which is currently running \cite{t2k} 
and \nova \cite{novareport} which is expected to start taking data in near future. 
Earlier studies on measurement of leptonic CP violation by 
conventional superbeam 
experiments can be found in 
\cite{BurguetCastell:2001ez,lindner,t2k_globes2,Mena:2004sa,Mena:2005ri}.

A potential problem in determining  $\dcp$ comes from 
the lack of knowledge of hierarchy which gives rise to  
wrong hierarchy-wrong $\dcp$ 
solutions 
\cite{Minakata:2001qm,degen_8fold}.
A prior knowledge of hierarchy can help to eliminate these
fake solutions thereby enhancing the CP sensitivity.  
However since the baselines of T2K and \nova are not too large 
they have limited hierarchy sensitivity.  Moreover this depends 
on the true  value  of $\dcp$  chosen by nature
\cite{Minakata:2001qm,Minakata:2003wq}.  
It has been shown recently in \cite{novat2k} 
that these experiments can determine hierarchy 
at 90\% C.L. only for favourable combination of
parameters --- (\{$\dcp \in [-180^\circ,0^\circ]$, NH\} or \{$\dcp \in [0^\circ,180^\circ]$, IH\}). However for the complementary unfavourable 
combinations, the hierarchy sensitivity of these experiments is low, because of 
which their $\dcp$ sensitivity is compromised.    

In this paper, we expand on the observation made in \cite{cpv_ino} regarding the
synergy between existing and upcoming atmospheric and long-baseline experiments
for measuring $\dcp$.
The central idea is that due to the large value of $\theta_{13}$  
atmospheric neutrinos passing through the earth experience 
appreciable matter effects leading to an enhanced hierarchy sensitivity. 
Moreover this sensitivity 
does not depend crucially on the true $\dcp$ values.
Thus addition of atmospheric information to the data from LBL experiments can
increase the hierarchy sensitivity in the  unfavourable region for the latter.
This feature leads to an  enhanced CP sensitivity for LBL experiments
when atmospheric data is included in the analysis \cite{cpv_ino}. 
This is despite atmospheric neutrino experiments themselves not having 
any appreciable sensitivity to $\dcp$ \cite{Gandhi:2007td}. 
Usually, studies of CP sensitivity are done assuming hierarchy 
and octant to be known. In our study we quantify  explicitly the exposures 
required by a realistic atmospheric neutrino experiment to 
achieve this.

We analyze in detail the individual and
combined $\dcp$ sensitivity of LBL  
and atmospheric neutrino
experiments, 
both at the level of oscillation probabilities and with simulations of relevant
experimental set-ups.
For the long-baseline experiments we consider T2K, which is already running 
and \nova which is expected to start taking data in 2014. For  
atmospheric neutrinos we choose the magnetized iron calorimeter detector
(ICAL) which is being constructed by the
India-based Neutrino Observatory (INO) collaboration  \footnote{
There are other upcoming atmospheric neutrino
experiments like Hyper-Kamiokande \cite{hkloi} and PINGU \cite{smirnov} 
which may provide
similar results}. 
For our study we adopt a staged approach where we look at the data which 
will be
available to us from these experiments at different chronological points 
over the next 15 years.   
We explore whether the LBL experiments T2K and \nova can give any evidence of   
$\dcp$ being different from 0 and $180^\circ$ by themselves or in combination with
data from ICAL@INO. 
We also present  results for the precision measurement of $\dcp$  
individually for the LBL 
experiments and in combination with ICAL@INO.
In addition to the role of atmospheric neutrino data in 
resolving hierarchy-$\dcp$ degeneracy, its impact on removing  
octant-$\dcp$ degeneracy is also examined. 
We also study how the CP sensitivity of the LBL experiments 
varies with $\theta_{13}$ in its current range. 

Many future experiments are being planned for addressing  the problem 
of resolution of mass hierarchy and determination of the CP phase $\dcp$. 
This includes LBNE
\cite{lbne}, LBNO \cite{lbno}, T2HK  \cite{t2hk}, ESS \cite{ESS} etc.  
The planning of these facilities are expected to  benefit
from a detailed assessment of the capabilities of the
currently running or under
construction experiments \cite{ourlbnopaper}. 
In this context it also makes sense to survey if T2K and \nova do not 
see CP violation with their currently projected run times then whether they 
can achieve this with extended run times \cite{novat2k,coloma,minakata2}. 
In this paper  we expound this   
possibility to explore 
the ultimate
reach of these experiments to detect $\dcp$.  

The paper is organized as follows. In Section II, we describe the $\dcp$
dependence of neutrino oscillation probabilities and how it is correlated
with other 
parameters. 
Section III gives the experimental details of the long-baseline experiments
(\nova and T2K) and atmospheric neutrino experiment (ICAL) considered in our
study. Section IV outlines the results for the $\dcp$ measurement and CP
violation discovery potential of \nova and T2K with different exposures
corresponding to different points of time in the future. 
{{In Section V we discuss the dependence of the CP sensitivity of \nova 
and T2K on neutrino parameters and the synergies between different channels. 
Section VI analyzes 
the CP sensitivity of atmospheric neutrino experiments with a magnetized iron
detector, focusing on 
the CP measurement potential for a combination of \nova and T2K
with ICAL}}. 
We summarize the
conclusions in Section VII.

\section{Effect of hierarchy and Octant on $\dcp$ sensitivity}

The sensitivity to $\dcp$ and potential for CP violation discovery can be
understood 
from the oscillation probabilities in matter
\cite{kimura,
ohlsson_acp,smirnov}.
The predominant contribution to
the $\dcp$ sensitivity is from 
the $\nu_{\mu} \rightarrow \nu_{e}$ oscillation probability ($\pmue$),
which has a dependence on $\dcp$ in its sub-leading term which is suppressed by
the small solar mass-squared difference. In matter of constant 
density, $\pmue$ can be expressed in terms
of the small parameters $\alpha = \Delta_{21}/\Delta_{31}$ and $s_{13}$
as \cite{akhmedov,cervera,freund}
\bea
P_{\mu e }&=
&4 s_{13}^2 s_{23}^2 \frac{\sin^2{[(1 -\hat A)\Delta]}}{(1-\hat A)^2}
\nn \\
&&
+ \alpha \sin{2\theta_{13}}  \sin{2\theta_{12}} \sin{2\theta_{23}} \cos{(\Delta
+ \dcp)}
\times \nn \\
&&
\frac{\sin{\hat A \Delta}}{\hat A} \frac{\sin{[(1-\hat A)\Delta]}}{(1-\hat A)}
+ {\cal{O}}(\alpha^2) 
\label{P-mue}
\eea
Here $\Delta_{ij} = m_i^2 - m_j^2$, $\Delta = \Delta_{31}L/4E$, 
$s_{ij} (c_{ij}) \equiv \sin{\theta_{ij}}(\cos{\theta_{ij}})$ and
$\hat{A} = 2\sqrt{2} G_F n_e E / \Delta_{31}$.
$G_F$ is the Fermi constant and $n_e$ is the electron number density.
The behaviour of this expression depends on the neutrino mass hierarchy, i.e.
the sign of the atmospheric mass-squared difference $\Delta_{31}$.
For neutrinos, $\hat{A}$ is positive for NH 
and negative for IH, while for antineutrinos it is the
opposite. 
It can be seen that the $\dcp$ dependence of the second sub-leading term
comprises
of both $\sin \dcp$ and $\cos \dcp$, and has the potential for discovering CP
violation. 

From Eq.\ref{P-mue}, the following observations can be made:

\begin{figure}[hbt]
\hspace{-0.2in}
\epsfig{file=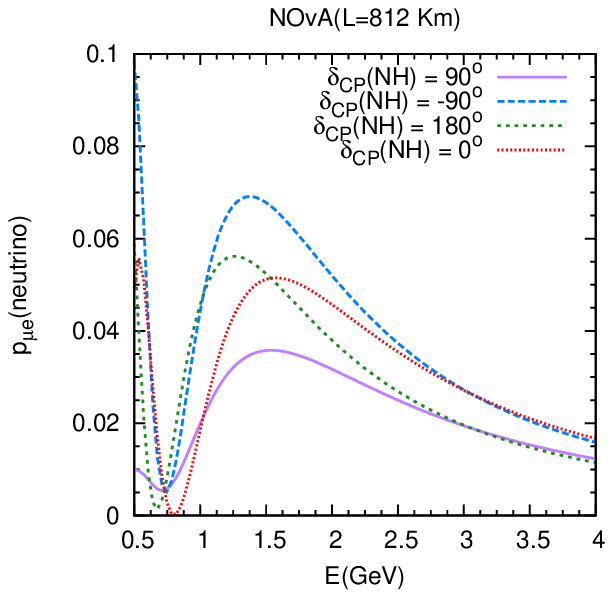, width=0.45\textwidth, bbllx=80, bblly=50, bburx=260, bbury=235,clip=}
\epsfig{file=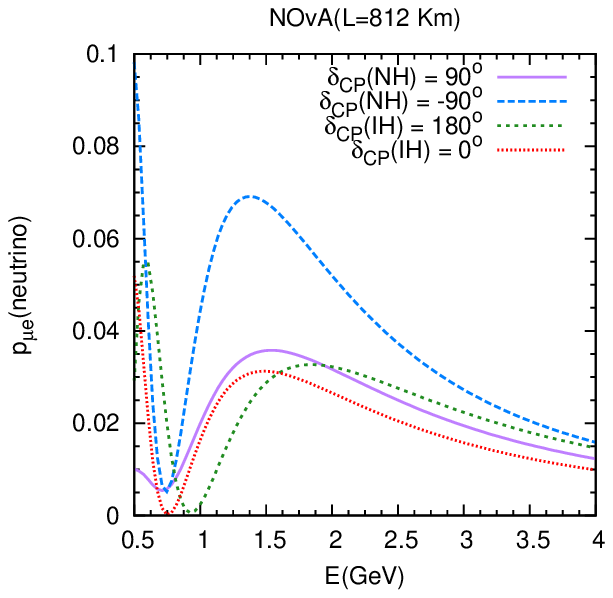, width=0.45\textwidth, bbllx=80, bblly=50, bburx=260, bbury=235,clip=}
\caption{{$\pmue$ energy spectrum for true $\dcp=90^\circ$ and $-90^\circ$ for true
NH 
(both panels), test $\dcp=0$ and $180^\circ$ for NH (left panel) and IH (right panel).
Here $\theta_{23}=39^\circ$, $\sin^2 2\theta_{13} = 0.1$.
}}
\label{Pdcp}
\end{figure}

\begin{figure}[hbt]
\hspace{-0.2in}
\epsfig{file=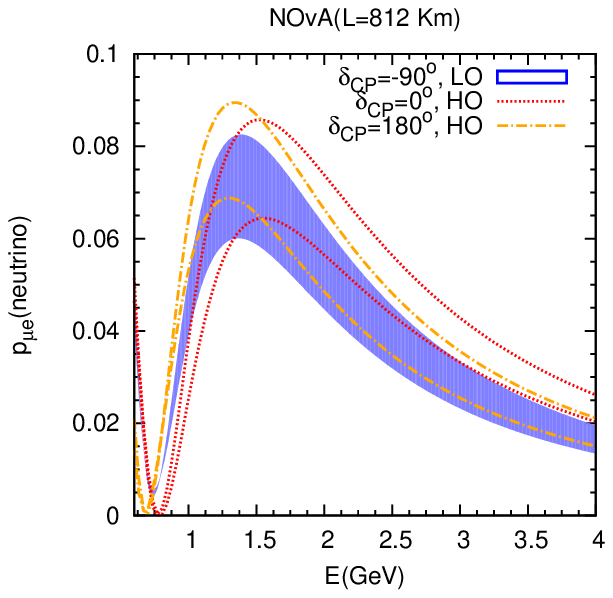, width=0.45\textwidth, bbllx=80, bblly=50, bburx=260, bbury=235,clip=}
\epsfig{file=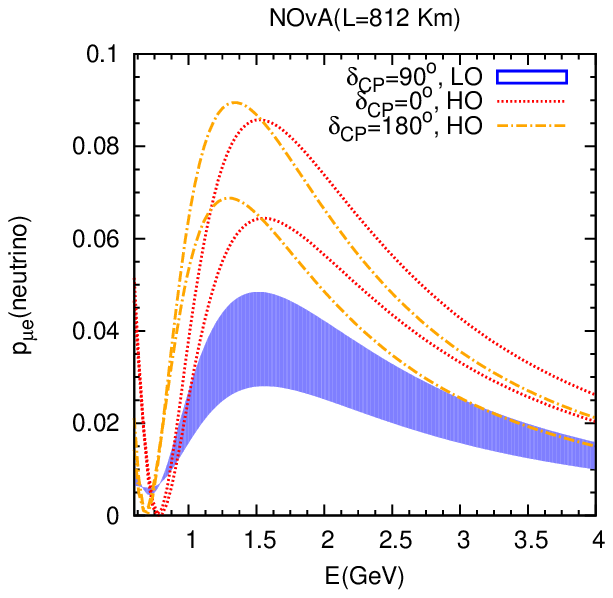, width=0.45\textwidth, bbllx=80, bblly=50, bburx=260, bbury=235,clip=}
\vspace{0.2in}
\epsfig{file=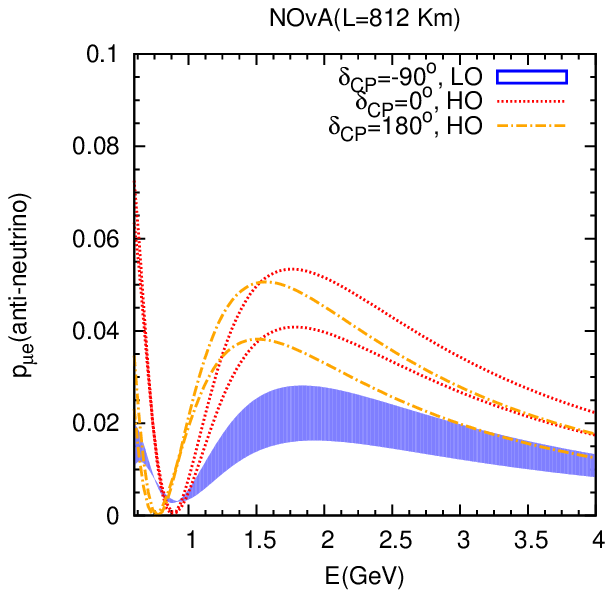, width=0.45\textwidth, bbllx=80, bblly=50, bburx=260, bbury=235,clip=}
\epsfig{file=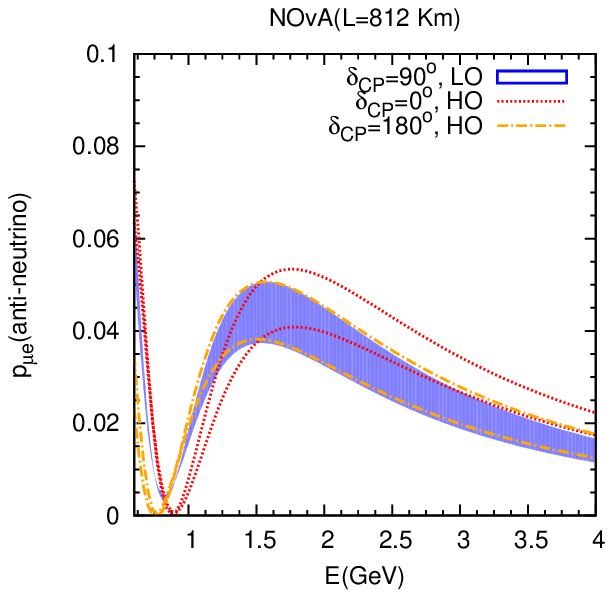, width=0.45\textwidth, bbllx=80, bblly=50, bburx=260, bbury=235,clip=}
\caption{{$\pmue$ energy spectrum showing the octant-$\dcp$ degeneracy. 
The upper panels are for neutrinos  whereas the lower panels are for 
antineutrinos.  
}}
\label{octant-dcp}
\end{figure}

\begin{enumerate}

\item Since $\dcp$ appears in the expression coupled with the atmospheric
mass-squared 
difference in the term $\cos(\Delta + \dcp)$,
it suffers from the hierarchy-$\dcp$ degeneracy which potentially limits the
CP sensitivity \cite{degen_8fold}.
The ambiguity shows up only in specific half-planes of true $\dcp$ depending on
the 
true mass hierarchy \cite{novat2k}. For neutrinos, the probability $\pmue$
is higher for NH than for IH due to matter effects as seen from the first term 
in Eq.~\ref{P-mue}. 
Near the oscillation maximum ($\Delta=90^\circ$), this increase in $\pmue$ 
can be compensated by the second term, if $\dcp$ lies close to $90^\circ$. 
Similarly, the lowering of $\pmue$ in case of IH can be compensated if 
$\dcp$ lies close to $-90^\circ$. To summarize, it is difficult to disentangle 
the effects of hierarchy and $\dcp$ if the hierarchy is normal and $\dcp$ lies in 
the upper half-plane (UHP,$\dcp \in (0,180^\circ)$) or if the hierarchy is inverted 
and $\dcp$ lies in the lower half-plane (LHP,$\dcp \in (-180^\circ,0)$). 
These unfavourable combinations of parameters make the NH and IH probabilities 
overlap, making hierarchy determination difficult.
On the other hand, 
if the combination of hierarchy and $\dcp$ in nature is either \{NH, LHP\} or 
\{IH,UHP\}, then there is a substantial separation between NH and IH and it is 
easier to distinguish between the two hierarchies \cite{novat2k}. 

 We illustrate this point with the help of Fig.\ref{Pdcp}. In this figure, 
$\pmue$ is plotted for the \nova 
baseline (812 Km) with the true values $\dcp=90^\circ$ and $-90^\circ$ 
and true NH (in both panels). The test curves in the left(right) panel are 
for $\dcp=0$ and $180^\circ$ for NH(IH). 
Thus, the left(right) plots show the separation between 
CP-conserving test $\dcp$ and maximally CP-violating true $\dcp$ when the 
hierarchy is known(unknown). This 
is indicative of the CP violation discovery potential in each case.
We see that the separations of the cases $\{ NH,90^\circ \}$ and
$\{ NH,-90^\circ\}$ from the CP conserving cases 
$\{ NH,0 \textrm{ or } 180^\circ\}$ are comparable. But if the test hierarchy 
for the CP-conserving curves is inverted, the separation is considerably
increased for 
$\dcp = -90^\circ$ and reduced for $\dcp = 90^\circ$. 
This indicates that a true value of $\dcp$ in the LHP
would be effective for lifting the $\dcp$-hierarchy degeneracy for 
true NH-test IH
while 
true $\dcp$ lying in the UHP would worsen the degeneracy. 
For antineutrinos and test hierarchy NH, 
since the matter effect is less the 
probability will be somewhat lower than the corresponding
ones for the neutrinos. In addition 
the curves for $+90^\circ$ and $-90^\circ$ will be interchanged  
but the separation for these two cases from the CP conserving values 
still remain comparable. 
When the test hierarchy is  IH, the antineutrino probabilities are higher 
because of enhanced matter effects. This compounded with the flipping of 
the $+90^\circ$ and $-90^\circ$ probability curves for NH leads to 
LHP still being favorable for lifting the degeneracy. 
For true IH, the opposite is true, i.e. $\dcp$ values in the UHP are favourable 
for resolving the degeneracy for both neutrinos and antineutrinos.

\item Since the $\dcp$-sensitive term occurs with $\sin 2\theta_{23}$, it 
gives rise 
to the intrinsic octant  degeneracy: 
$P(\theta_{23}) = P (90^\circ - \theta_{23})$ \cite{Fogli:1996pv}.
However recently 
a more generalized form of the octant degeneracy: 
$P(\theta_{23}^{tr},\theta_{13},\dcp) = P(\theta_{23}^{wrong},\theta_{13}',\dcp')$
has been  
elucidated in detail in \cite{usoctant,ourprl}, extending the conventional definition.
This includes the possibility that the test value of $\theta_{23}$ occurring 
anywhere  
in the 'wrong' octant may give the same probability. 
The recent tighter constraint on the value of $\theta_{13}$ helps to weaken the degeneracy with this parameter, but the ambiguity between the two octants for different values of $\dcp$ still remains.

Whether this degeneracy is manifested in the
results
for CP sensitivity from an experiment depends on whether
the experiment (or combination of experiments) is capable of determining the
octant and 
resolving the octant-$\dcp$ degeneracy to
a good enough confidence level. The octant sensitivity arises from the 
$\sin^2\theta_{23}$ dependence in the leading term, 
and is significantly dependent on 
the true values of oscillation parameters $\dcp$ and $\theta_{23}$
\cite{usoctant}. 
Since $\pmue$ increases with $\theta_{23}$ and with $\dcp$ in the LHP, while
$\dcp$ in the UHP pulls it down, a true value of $\theta_{23}$ lying in the
higher octant (HO) would be more likely to suffer from the octant-$\dcp$
degeneracy if the true value of $\dcp$ is in the UHP. On the other hand, if true
$\theta_{23}$ is in the lower octant (LO) then true $\dcp$ lying in the LHP
would raise $\pmue$ and lead to an ambiguity with  $\pmue$  values corresponding
to  test $\theta_{23}$ in the HO. Hence the LHP is favourable for resolving the
octant $\dcp$ degeneracy in the case of true HO and the UHP is favourable for
true LO.
These features are reflected in Fig. \ref{octant-dcp} where we 
show the effect of octant degeneracy  in distinguishing between the 
CP conserving and maximally CP violating cases for $\pmue$, for the 
\nova baseline. 
The upper panel is for neutrinos whereas the lower panel is for antineutrinos. 
The shaded region corresponds to true LO. The first panel shows that 
for true LO and  true $\dcp=-90^\circ$ (LHP) the two cases cannot be
distinguished whereas  in the second panel, for 
$\dcp= 90^\circ$ (UHP) a clear separation is seen.
For antineutrinos the behaviour for LHP and UHP is opposite.
This indicates that a combination of neutrinos and antineutrinos would 
be conducive for removal of octant-$\dcp$ degeneracy  \cite{uss,minakata2}. 
For true HO the behaviour is opposite.
\end{enumerate}

\section{Experimental details}

For the long-baseline experiments \nova and T2K, simulation is done using 
the GLoBES package~\cite{globes,globes2,globes_xsec,globes-xsec2}.
 T2K (L $=$ 295 Km) is assumed to have a $22.5$ kt Water \^{C}erenkov detector
and a $0.77$ MW beam 
running effectively for $5(\nu) + 0(\overline{\nu})$ or $3(\nu) +
2(\overline{\nu})$ years by 2016. 
The initial plan of T2K was to run with
with 10$^{21}$ protons on target (pot)/year for five years \cite{t2k_globes}. 
However because of natural calamity T2K has not been 
able to achieve its full capacity yet.  We take into account its present lower 
power run as well as the planned upgrades to give a total of 5 effective T2K 
years till 2016 (i.e. a total of $5 \times 10^{21}$ pot). 
We also consider the option of T2K running for $5(\nu) + 5(\overline{\nu})$
years by 2021, to ascertain whether such an extension would be advantageous. 
For \nova (L $=$ 812 Km), we consider a $14$ kt TASD detector 
with a $0.7$ MW beam with 7.3 $\times$ 10$^{20}$ pot/year running for 
$3(\nu) + 3(\overline{\nu})$ years by 2020 and $5(\nu) + 5(\overline{\nu})$
years by 2024. 
In this work we use a re-optimized \nova experimental set-up with refined event 
selection criteria \cite{nova_reopt,nova_reopt2}.
Detailed specifications of these experiments are given in 
\cite{t2k_globes,t2k_globes2,t2k_globes3,t2k_globes4,t2k_globes5,novareport,
nova_reopt2}.

For atmospheric neutrinos, we analyze 
a magnetized iron calorimeter detector (ICAL) of the prototype planned by the 
India-based Neutrino Observatory (INO),
which will detect muon events with the capability of charge identification
\cite{gct}. We use constant neutrino
energy and angular detector resolutions of 10$\%$ and 10$^\circ$ respectively, 
unless otherwise specified.   Note that the neutrino resolutions using 
INO simulation codes are currently being generated. However we have checked that 
the resolutions used above 
gives similar results as obtained by the INO simulation code using muons \cite{gct}. 
We consider a 1 GeV neutrino energy threshold, 85\% efficiency and 100\% charge 
identification efficiency. 
We look at two detector exposures of 250 kT yr, corresponding to 5 years of
running for a 50 kT detector, and 500 kT yr or 10 years of running with such a
detector. The detector is currently under construction, with a projected time
frame of 5 years to completion, so this data is expected to be available by
about 2023 and 2028 respectively. 
{{Earth matter effects are included in the atmospheric neutrino analysis using
a standard Preliminary Reference Earth Model (PREM) density profile of the earth \cite{prem}}}.

 Henceforth, we give the exposure of \nova or T2K as $a+b$ where $a$ and $b$
 respectively denote the number of years of neutrino and antineutrino running of
 the experiment.

For our study of T2K, \nova and ICAL, we look at the following chronological
points:

\begin{itemize}
\item 2016, when T2K will have completed either a (5+0) or a (3+2) run
\item 2020, when \nova will have completed a (3+3) run 
\item 2024, when \nova will have completed a (5+5) run and ICAL will have at least
5 years of data
\item 2028, when ICAL will have 10 years of data
\end{itemize}

We also consider the case of T2K going on to a (5+5) run, which can be taken into
account in the 2024 analysis along with \nova (5+5) and ICAL 5.

\section{CP sensitivity of T2K and \nova: chronological progression}

In this section, we study the prospects for CP violation discovery and 
$\dcp$ precision measurement of \nova and T2K for different exposures 
corresponding to
progressive points of time in the next 10 years. The experimental capabilities 
are demonstrated using CP violation discovery plots and $\dcp$ precision plots 
respectively.

The discovery potential of an experiment for CP violation is computed by 
considering a variation of the $\dcp$ over the full range $[0,180^\circ)$ in the 
simulated true or `experimental' event spectrum $N_{ex}$, and comparing this 
with $\dcp = 0$ or $180^\circ$ in the test or `theoretical' event spectrum $N_{th}$.  
The discovery $\chi^2$ in its simplest statistical form is defined as 
\footnote{In this work we have used the standard definition of $\chi^2$
and used the usual convention $N_\sigma = \sqrt{\Delta \chi^2}$.
Note that since we have not included fluctuations in the simulated data 
in our case $\Delta \chi^2 = \chi^2$. 
For alternate statistical analysis using
frequentist or Bayesian approach see 
\cite{Blennow:2013oma,Qian:2012zn,Ciuffoli:2013rz,Schwetz:2006md,Blennow:2013kga}.}
\be
\chi^2 = \text{min} \frac{(N_{ex} (\dcp^{tr}) - N_{th} (\dcp^{test}=0,180^\circ))^2}{N_{ex}
(\dcp^{tr})}
\label{chisq}
\ee
In our calculation we include a
marginalization over 
systematic errors and uncertainties for each experiment. 
\footnote{We have not used correlated systematics for various experiments in this study. 
However, such an analysis would impose additional constraints on the systematic 
parameters and serve to improve the results presented here.} The resultant $\chi^2$ 
from the various experiments are then added and finally marginalized 
(unless specified otherwise) over the parameters  
$\theta_{23}$, $\theta_{13}$, $|\Delta_{31}|$ and hierarchy  
in the test spectrum.  
As expected, the discovery potential of the experiments is zero for true 
$\dcp=0$ and $180^\circ$, while it is close to maximum at the maximally CP violating 
values $\dcp=\pm 90^\circ$.
{{ We use the following transformations relating the effective measured values of the 
atmospheric parameters $\dmmm$ and $\tmm$ to their natural values $\Delta_{31}$ 
and $\theta_{23}$ \cite{dm31_defn,dm31_defn2,th23_defn}:
\begin{equation}
 \sin\theta_{23} = \frac{\sin\tmm}{\cos\theta_{13}} \ ~,
\end{equation}
\begin{equation}
 \Delta_{31} = \dmmm + (\cos^2\theta_{12} - \cos\delta\sin\theta_{13}\sin2\theta_{12}
\tan\theta_{23})\Delta_{21}\ ~.
\end{equation}
The effective values $\dmmm$ and $\tmm$ correspond to
parameters measured by muon disappearance experiments. It is 
advocated to use these values in the definitions of priors if 
the prior is taken from muon disappearance measurements. 
The corrected definition of $\tmm$ is significant due to the large measured 
value
of $\theta_{13}$, while for $\dmmm$ the above transformation is valid 
even for 
small $\theta_{13}$ values.
In our analysis we do not use any external priors for these parameters 
as the experiments themselves are sensitive to these parameters. 
However it is to be noted that for the effective parameters, 
there is an exact mass hierarchy degeneracy between 
$\dmmm$ and $-\dmmm$ and an exact intrinsic 
octant degeneracy between $\theta_{\mu\mu}$ and $90^\circ - \theta_{\mu\mu}$. 
Therefore use of these values in the analysis ensures that one hits the exact 
minima for the wrong hierarchy and wrong octant in the numerical analysis 
for the muon disappearance channel. 
Measurements with the appearance channel and  
the presence of matter effects can break these degeneracies. 
Also, the generalized octant degeneracy occurring between values of 
$\theta_{\mu\mu}$ in opposite octants 
for different values of $\theta_{13}$ and $\dcp$ is still present for the 
effective atmospheric mixing angle. 
For such cases, a fine marginalization grid has to be 
used in the analysis in order to capture the $\chi^2$ minima occurring in the
wrong hierarchy and wrong octant. 

The following true values and test ranges of parameters are used in our computation:
\begin{eqnarray}
(\Delta_{21})^{tr} & = & 7.6 \times 10^{-5} \textrm{ eV}^2 \nonumber \\
(\sin^2\theta_{12})^{tr} & = & 0.31 \nonumber \\
(\sin^2 2\theta_{13})^{tr} & = & 0.1 \\
\dmmm^{tr} & = & 2.4 \times 10^{-3} \textrm{ eV}^2 \nonumber ~,
\label{eq:truevals}
\end{eqnarray}
with specific values
of $\tmm^{tr}$ and $\delta_{CP}^{tr}$.
\begin{eqnarray}
\theta_{\mu \mu}^{test} & \in & (35^o,55^o) \nonumber \\ 
\sin^2 2\theta_{13}^{test} & \in & (0.085,0.115) ~. \nonumber \\
\dmmm^{test} & \in & (2.2,2.6) \times 10^{-3} \textrm{ eV}^2 \nonumber \\
\label{eq:rangevals}
\end{eqnarray}
$\Delta_{21}$ and $\sin^2\theta_{12}$ are fixed to their true values
since their effect is negligible.
External (projected) information on $\theta_{13}$ from the reactor experiments 
is added in the form of a prior on $\theta_{13}$:   
\begin{equation}
\chi^2_{prior} = 
\left(\frac{{\sin^2 2\theta_{13}^{{tr}}} - \sin^2 2\theta_{13}}
{\sigma(\sin^2 2\theta_{13})}\right)^2
\end{equation}
with the 1$\sigma$ error range $\sigma_{\sin^2 2\theta_{13}} = 0.005$}}.
 
The CP sensitivity of an experiment can also be quantified by the precision
measurement of $\dcp$ 
possible by the experiment.
In this case we look at a variation of the $\dcp$ over the full range $[0,360^\circ)$
in both the 
simulated true event spectrum $N_{ex}$ and the test event spectrum $N_{th}$. 
Thus the precision $\chi^2$ is given by  
\be
\chi^2 = \text{min} \frac{(N_{ex} (\dcp^{tr}) - N_{th} (\dcp^{test}))^2}{N_{ex}
(\dcp^{tr})}
\label{chisq_precision}
\ee
We present precision plots which show the test $\dcp$ range allowed by the data for
each true value of $\dcp$, up to a specified confidence level. 
The allowed values of $\dcp$ are represented by the shaded regions 
in the figures. For an ideal
measurement, the allowed values would 
be very close to the true value. 
Thus the allowed region would be along the ${\dcp}^{tr} = {\dcp}^{test}$ diagonal. 
However, due to finite precision of the parameters as well as
the parameter degeneracies, other $\dcp$ values are also seen to be allowed.

\subsection{CP sensitivity of T2K (3+2) and (5+0) (2016)} 

T2K is expected to have a neutrino run of 5 years. There are 
also discussions for a break-up of neutrino and antineutrino runs, for which 
we consider the case of (3+2) years \cite{minakata2}. 
In the left panels of Fig.~\ref{t2k3250nova33}, we depict the CP violation discovery 
for both these options (upper row) and
90$\%$ C.L. $\dcp$ precision for T2K (3+2) (middle row) and T2K (5+0)
(bottom row) for $\theta_{\mu\mu}=39^o$, $\sin^2 2\theta_{13} = 0.1$ (true values) 
and true NH.
The figure shows that the CP sensitivity of T2K alone is quite low, especially
for the (5+0) case, where the discovery potential remains below 
$\chi^2 = 2$ over the entire true $\dcp$ range. This is because the baseline of
T2K (295 Km) is relatively short and earth matter effects are minimal, leading
to the hierarchy-$\dcp$ degeneracy predominating in both half-planes when only a
neutrino beam is taken. When we consider a neutrino-antineutrino combination,
the { different} behaviours of the neutrino and antineutrino probabilities
partially resolves the degeneracy in the favourable half-plane (in this case the
LHP) for a (3+2) run. Therefore, as pointed out in \cite{minakata2}, a T2K (3+2) run
provides better CP sensitivity than a T2K (5+0) run. This is also evident in
the precision plots, where the allowed region of $\dcp$ (shaded area) is more
for the (5+0) case, indicating that less regions of $\dcp$ are excluded at 90$\%$
C.L.

\subsection{CP sensitivity of T2K (3+2) and (5+0) with \nova (3+3) (2020)}

\begin{figure}
\hspace{-0.2in}
\epsfig{file=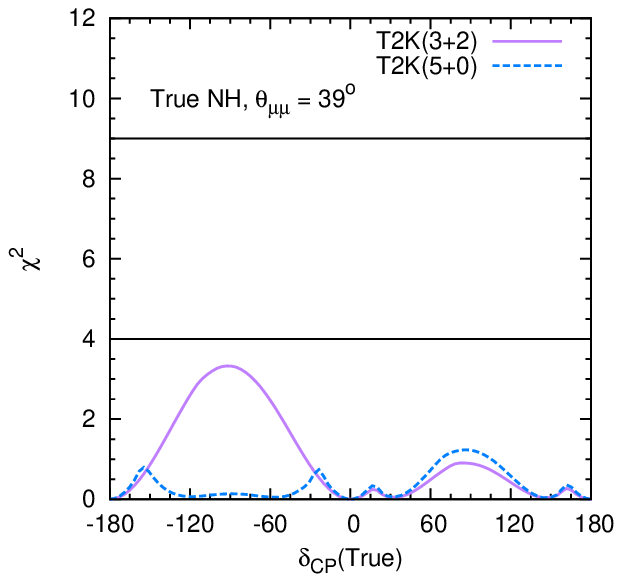, width=0.45\textwidth, bbllx=80, bblly=50, bburx=265, bbury=235,clip=}
\epsfig{file=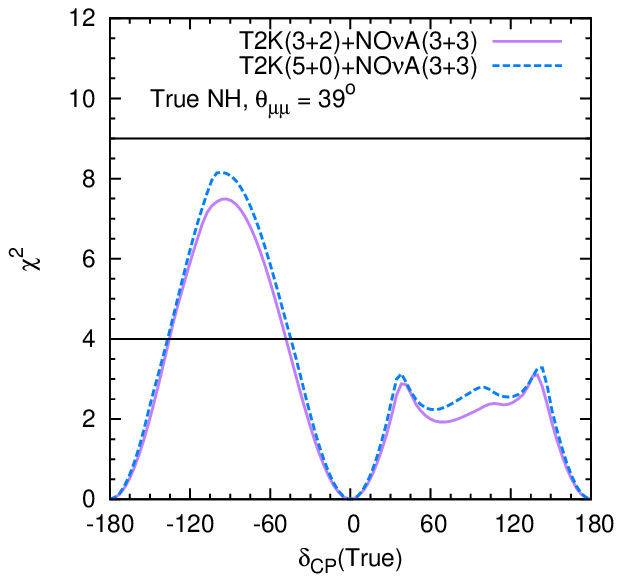, width=0.45\textwidth, bbllx=80, bblly=50, bburx=265, bbury=235,clip=} \\
\epsfig{file=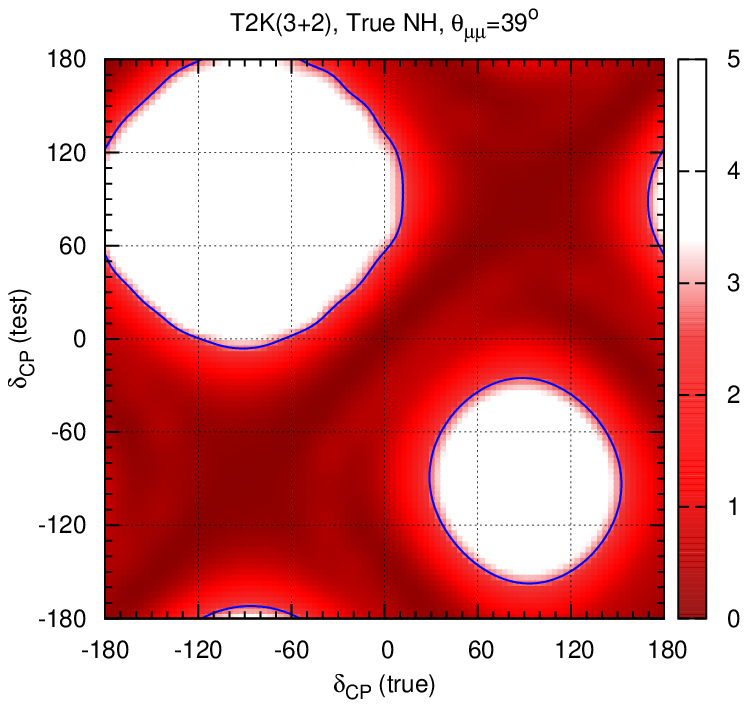, width=0.45\textwidth, bbllx=120, bblly=70, bburx=335, bbury=270,clip=}
\epsfig{file=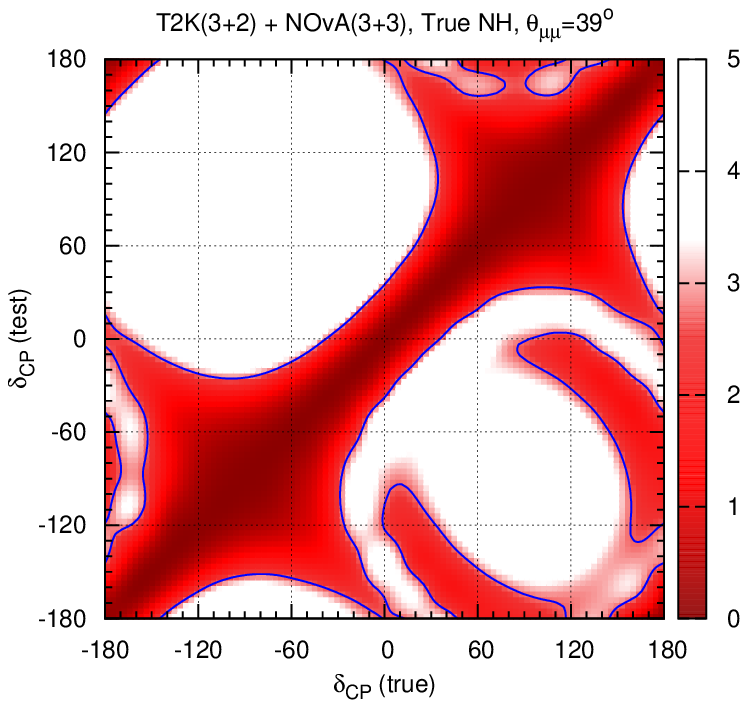, width=0.45\textwidth, bbllx=120, bblly=70, bburx=335, bbury=270,clip=} \\
\epsfig{file=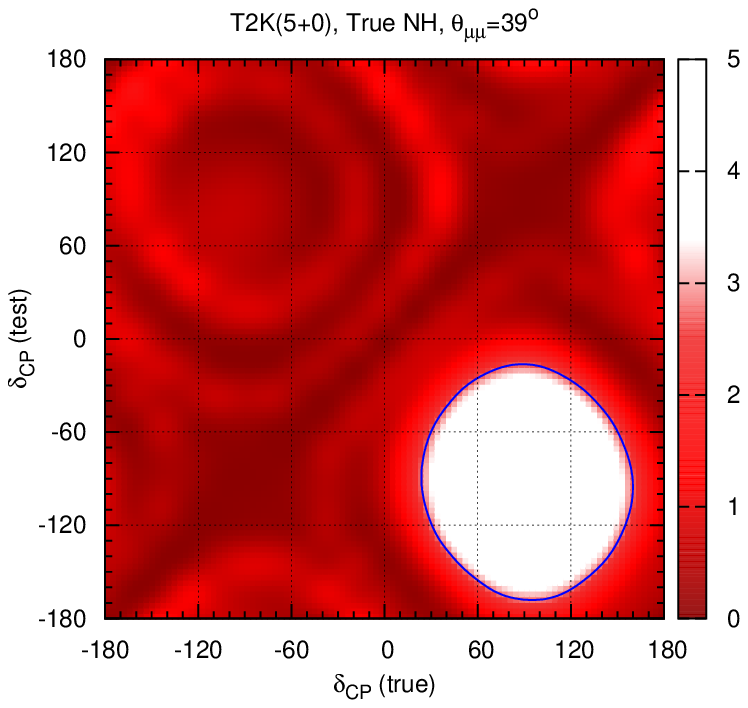, width=0.45\textwidth, bbllx=120, bblly=70, bburx=335, bbury=270,clip=}
\epsfig{file=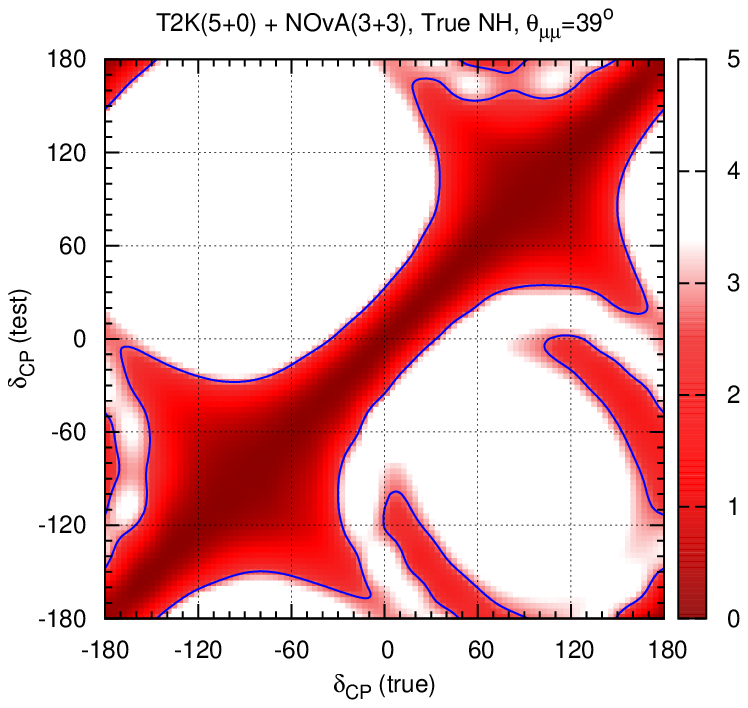, width=0.45\textwidth, bbllx=120, bblly=70, bburx=335, bbury=270,clip=}
\caption{{CP violation discovery (upper row) and 90$\%$ C.L. $\dcp$ precision
(middle and lower rows) for T2K (left panels) and T2K + \nova (right panels)
for $\theta_{\mu\mu}=39^o$, $\sin^2 2\theta_{13} = 0.1$ and true NH. 
}}
\label{t2k3250nova33}
\end{figure}

The experiments T2K and \nova are synergistic since the different baselines (295
Km for T2K and 812 Km for \nova) experience
different degrees of earth matter effects and hence show somewhat different
dependences on the neutrino parameters.
In particular, the degeneracies observed in Fig.~\ref{t2k3250nova33} can be resolved in
some areas by combining T2K with \nova.
We explore how addition of \nova affects the difference in CP sensitivity between T2K
(3+2) and (5+0) runs.  

The right panels of Fig.~\ref{t2k3250nova33} shows the CP violation discovery 
(upper row) and 90$\%$
C.L. $\dcp$ precision (middle and lower rows) for T2K + \nova
for true NH and the same values of parameters as the left panels.
The upper row depicts the discovery potential for 
the combination of both T2K exposures with \nova (3+3).
The middle and lower rows show the $\dcp$ precision for T2K (3+2) + \nova (3+3) 
and T2K (5+0) + \nova (3+3) respectively.
A comparison of the left and right panels tells us that for both discovery and
precision, the advantage offered by T2K (3+2) over (5+0) is lost when
we combine T2K with \nova. While for T2K alone the discovery $\chi^2$ can rise
above 2 in the LHP for (3+2) but remains well below 2 for (5+0),
the discovery $\chi^2$ of \nova + T2K is nearly identical for T2K (3+2) and (5+0),
and rises to values above $\chi^2=6$ (2.5$\sigma$) in the LHP.
The allowed regions also look similar in the two cases. 

This behaviour can be explained as follows. Since \nova already includes a
combined neutrino-antineutrino run, it is 
capable of resolving the hierarchy-$\dcp$ degeneracy and providing significant
CP sensitivity in the favourable half-plane. Therefore the hierarchy degeneracy
resolution provided by T2K (3+2) in the favourable half-plane is no longer
required when T2K is combined with \nova. Thus in the combined analysis, the T2K
CP sensitivity adds to the \nova sensitivity irrespective of whether T2K has a
(5+0) or (3+2) run. 
For the subsequent chronological analysis, we choose the T2K run to comprise of (5+0) years.

\subsection{CP sensitivity of T2K (5+0) with \nova (5+5) (2024)}

\begin{figure}[hbt]
\hspace{-0.2in}
\epsfig{file=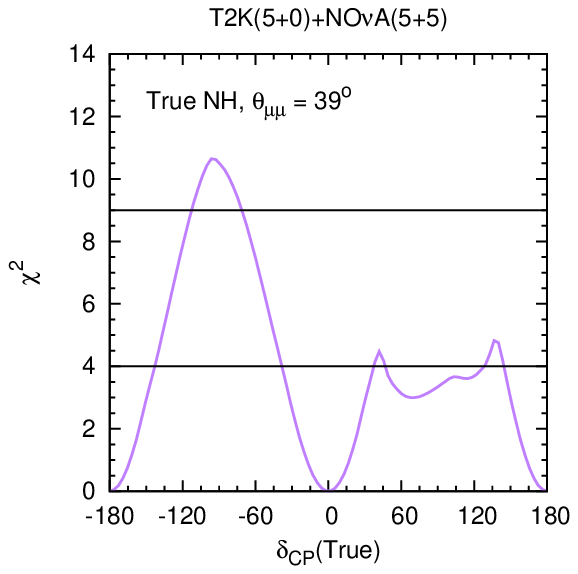, width=0.45\textwidth, bbllx=80, bblly=50, bburx=265, bbury=235,clip=}
\epsfig{file=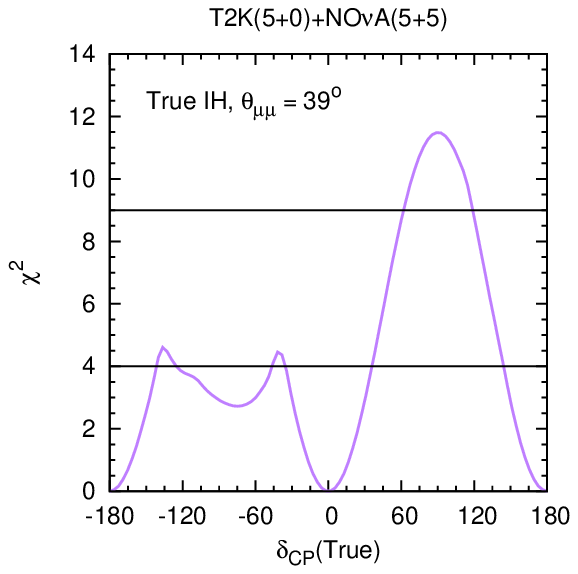, width=0.45\textwidth, bbllx=80, bblly=50, bburx=265, bbury=235,clip=} \\
\vspace{0.2in}
\epsfig{file=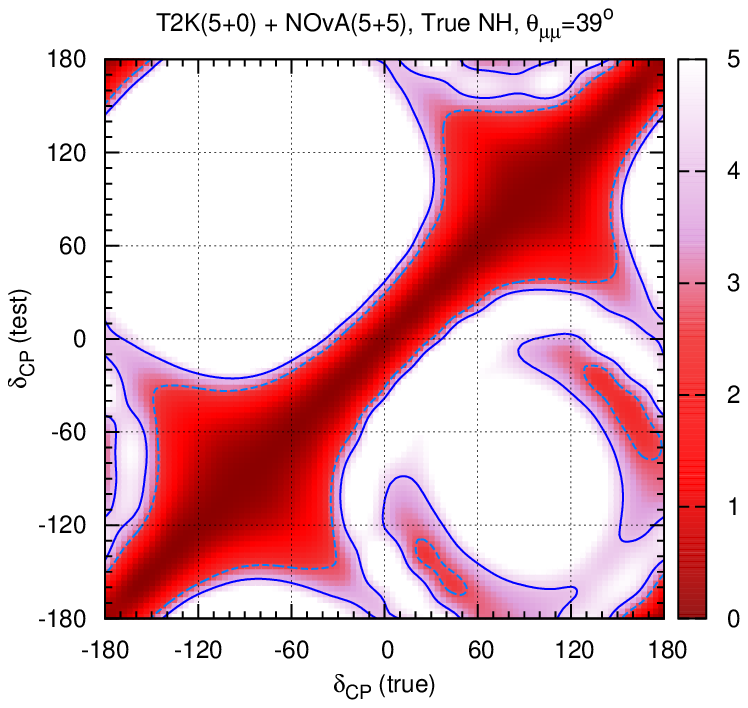, width=0.45\textwidth, bbllx=120, bblly=70, bburx=335, bbury=270,clip=}
\epsfig{file=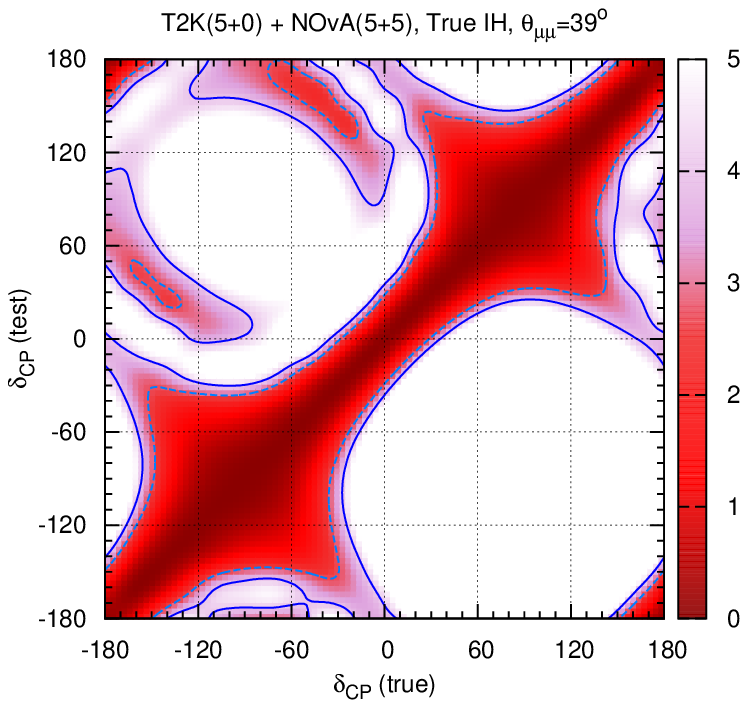, width=0.45\textwidth, bbllx=120, bblly=70, bburx=335, bbury=270,clip=}
\caption{{CP violation discovery (upper row) and 90$\%$/95$\%$ C.L. $\dcp$ 
precision (lower row) for \nova (5+5) +T2K (5+0) 
for $\theta_{\mu\mu}=39^\circ$, $\sin^2 2\theta_{13} = 0.1$ and true NH (left panel)
or 
IH (right panel).
}}
\label{t2k50nova55}
\end{figure}

Although the current projection of \nova is to run for (3+3) years we 
also consider the possibility of a (5+5) run of \nova. This is to investigate the 
possibility of an enhanced sensitivity to   $\dcp$
using upgradation of current facilities.   
In Fig.\ref{t2k50nova55}, we plot the CP violation discovery (upper row) and
90$\%$/95$\%$ C.L. $\dcp$ precision (lower row) for true NH (left panel) or
true IH (right panel). Comparing with
Fig.\ref{t2k3250nova33}, it can be observed that the increased \nova exposure
adds to the discovery potential, giving values as high as $\chi^2=9$ (3$\sigma$)
for maximal CP violation in the favourable half-plane in each case and reaching
close to $\chi^2=4$ (2$\sigma$) at some points in the unfavourable half-plane
even though the discovery minima still lie in the wrong-hierarchy region there. 
In the precision figures, the allowed regions shrink to  an area along the major
diagonal (true $\dcp$ = test $\dcp$) corresponding to the right-hierarchy solutions
and some off-axis islands corresponding to the wrong-hierarchy solutions
arising from the hierarchy-$\dcp$ degeneracy.
These are, as expected, in the UHP for true NH and in the LHP for true IH. 

\subsection{CP sensitivity of T2K (5+5) with \nova (5+5) (2024, alternative T2K
run)}

\begin{figure}[hbt]
\hspace{-0.2in}
\epsfig{file=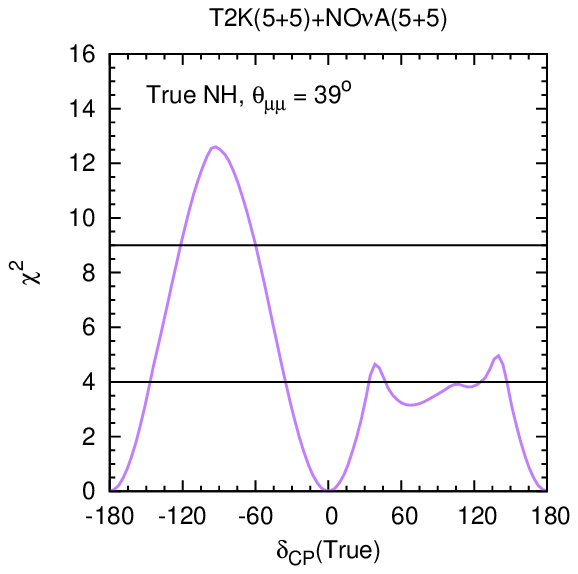, width=0.45\textwidth, bbllx=80, bblly=50, bburx=265, bbury=235,clip=}
\epsfig{file=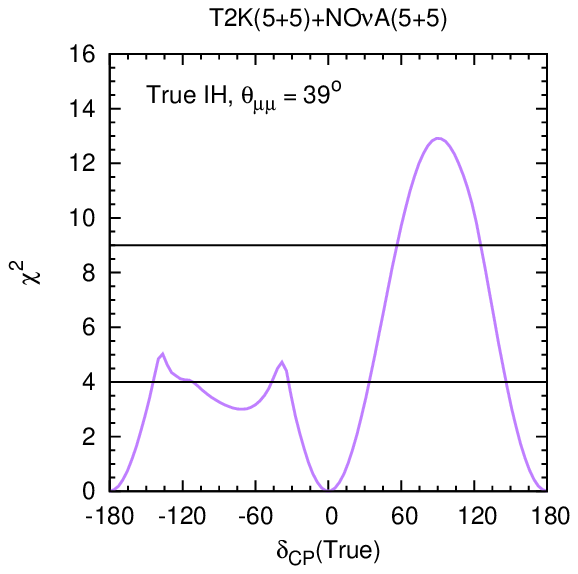, width=0.45\textwidth, bbllx=80, bblly=50, bburx=265, bbury=235,clip=} \\
\vspace{0.2in}
\epsfig{file=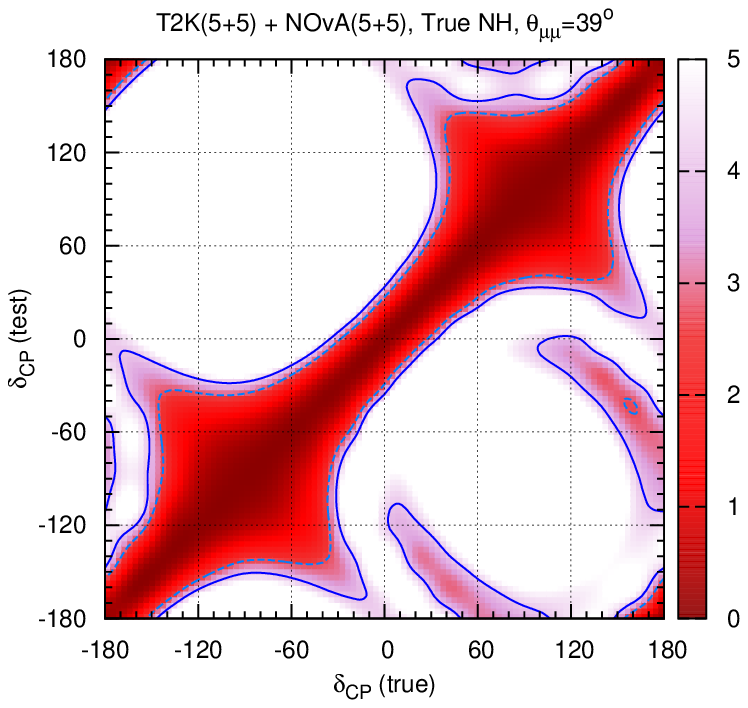, width=0.45\textwidth, bbllx=120, bblly=70, bburx=335, bbury=270,clip=}
\epsfig{file=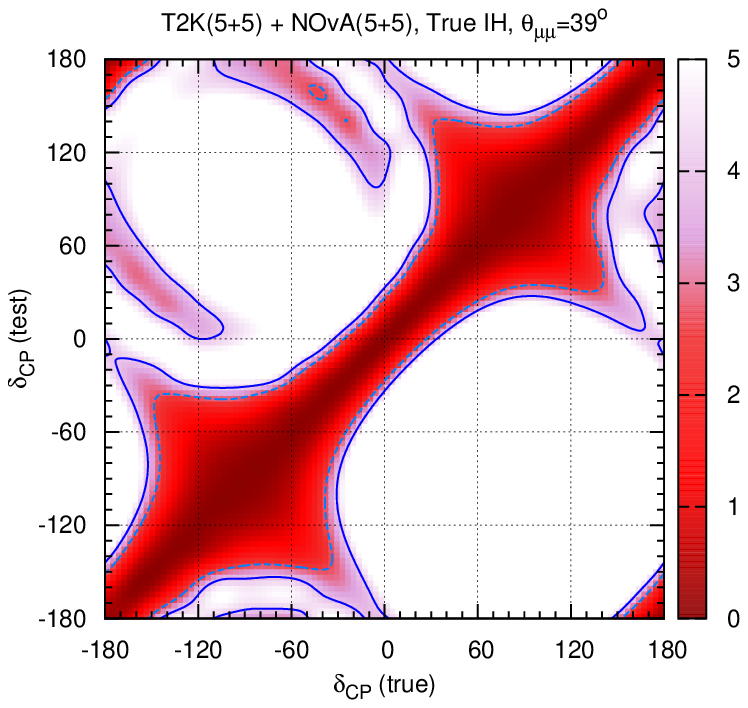, width=0.45\textwidth, bbllx=120, bblly=70, bburx=335, bbury=270,clip=}
\caption{{CP violation discovery (upper row) and 90$\%$/95$\%$ C.L. $\dcp$ precision
(lower row) for \nova (5+5) +T2K
(5+5) 
for $\theta_{\mu\mu}=39^\circ$, $\sin^2 2\theta_{13} = 0.1$ and true NH (left panel)
or 
IH (right panel).
}}
\label{t2k55nova55}
\end{figure}

\begin{figure}[hbt]
\hspace{-0.1in}
\epsfig{file=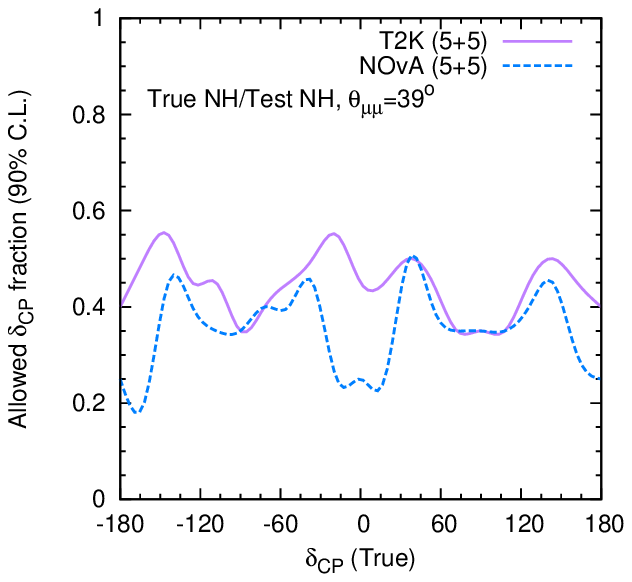, width=0.35\textwidth, bbllx=80, bblly=50, bburx=265, bbury=235,clip=}
\epsfig{file=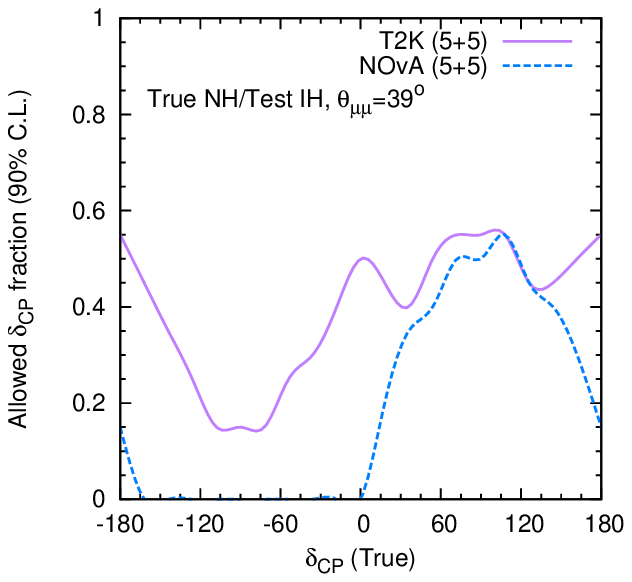, width=0.312\textwidth, height=5.8cm, bbllx=100, bblly=50, bburx=265, bbury=235,clip=} 
\epsfig{file=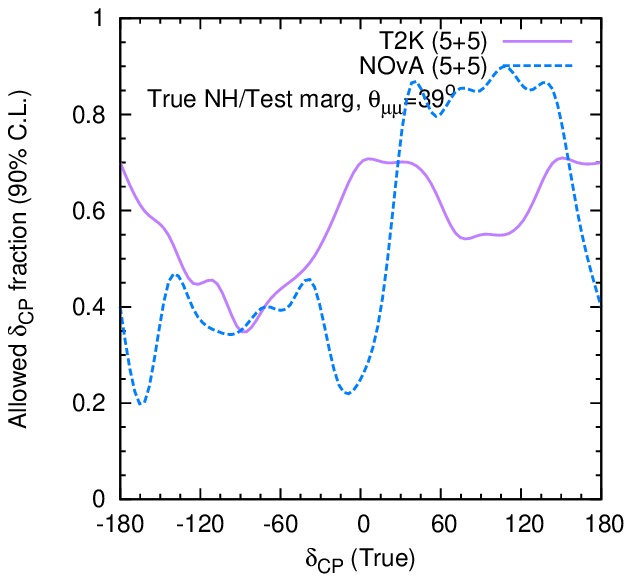, width=0.312\textwidth, height=5.8cm, bbllx=100, bblly=50, bburx=265, bbury=235,clip=} 
\caption{{Allowed CP fraction (90$\%$ C.L.) corresponding to each 
true $\dcp$ for \nova (5+5) and T2K (5+5) 
for $\theta_{\mu\mu}=39^\circ$, $\sin^2 2\theta_{13} = 0.1$ and true NH, with test
NH (left panel), test IH (middle panel)
and marginalization over hierarchy (right panel).
}}
\label{t2k55nova55cpfraction}
\end{figure}

\begin{figure}
\epsfig{file=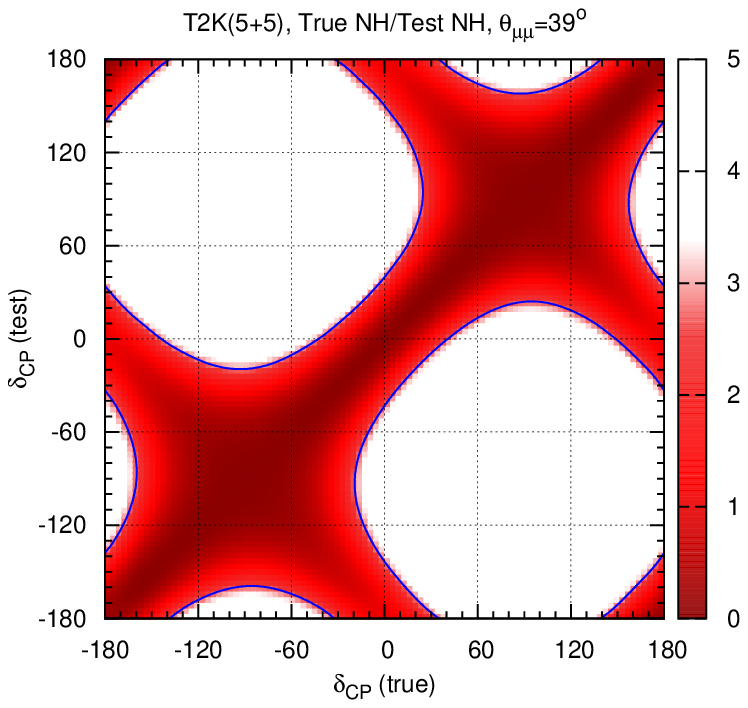, width=0.45\textwidth, bbllx=120, bblly=70, bburx=335, bbury=270,clip=}
\epsfig{file=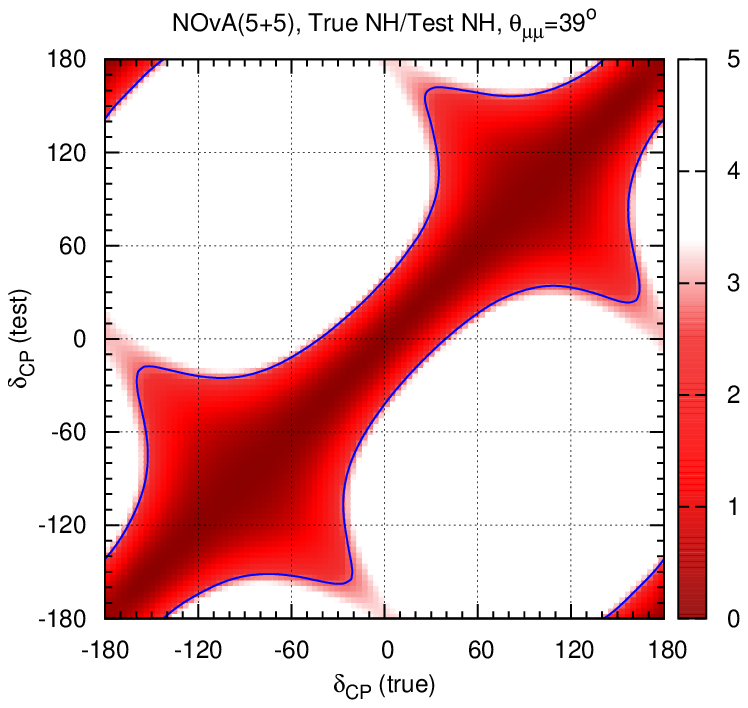, width=0.45\textwidth, bbllx=120, bblly=70, bburx=335, bbury=270,clip=}

\epsfig{file=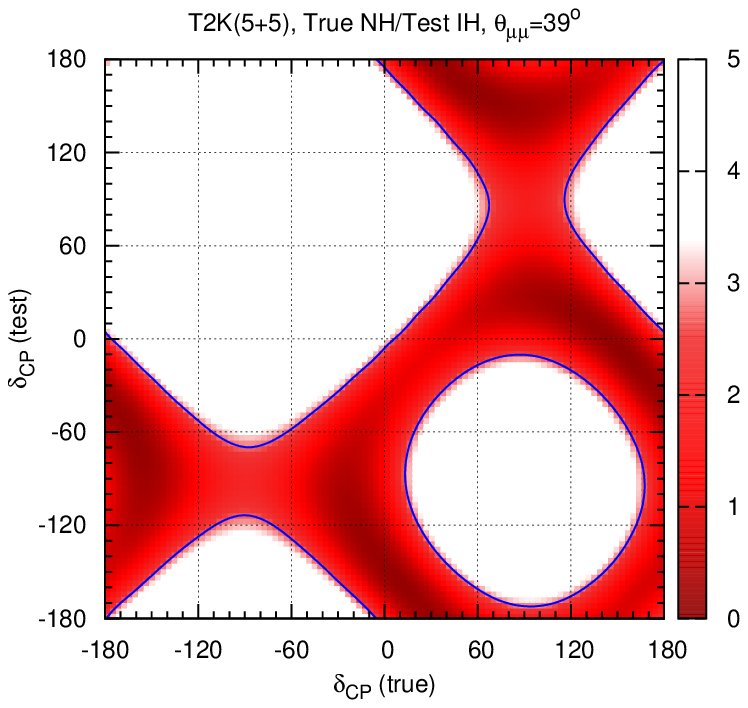, width=0.45\textwidth, bbllx=120, bblly=70, bburx=335, bbury=270,clip=}
\epsfig{file=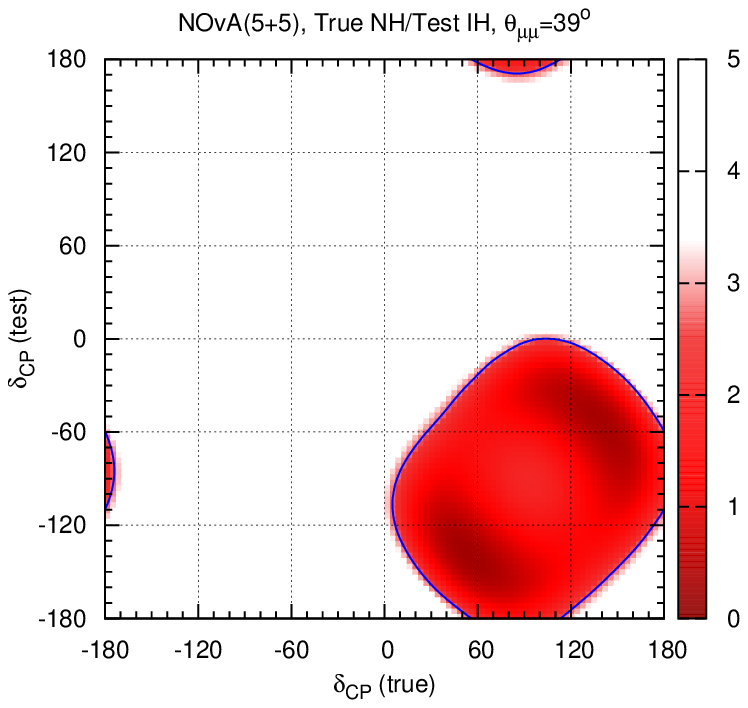, width=0.45\textwidth, bbllx=120, bblly=70, bburx=335, bbury=270,clip=}

\epsfig{file=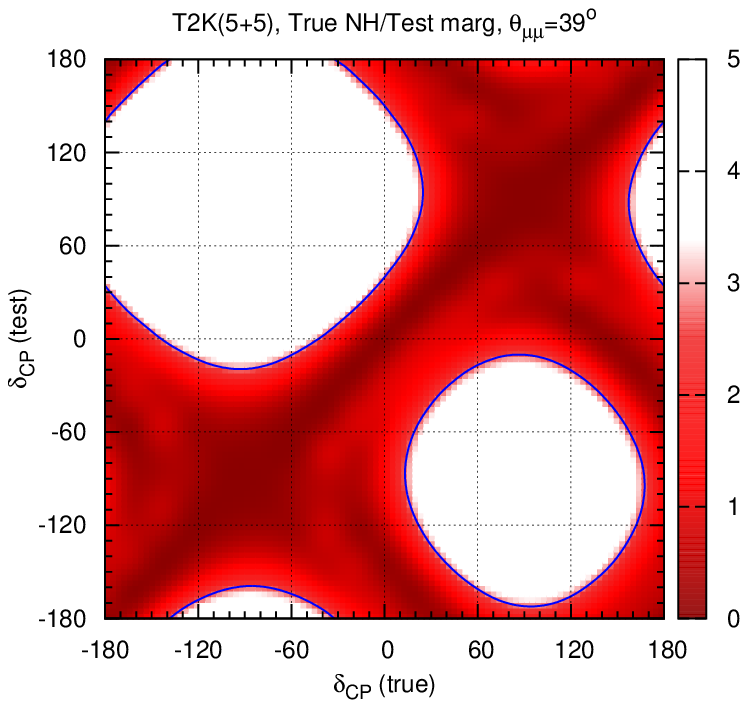, width=0.45\textwidth, bbllx=120, bblly=70, bburx=335, bbury=270,clip=}
\epsfig{file=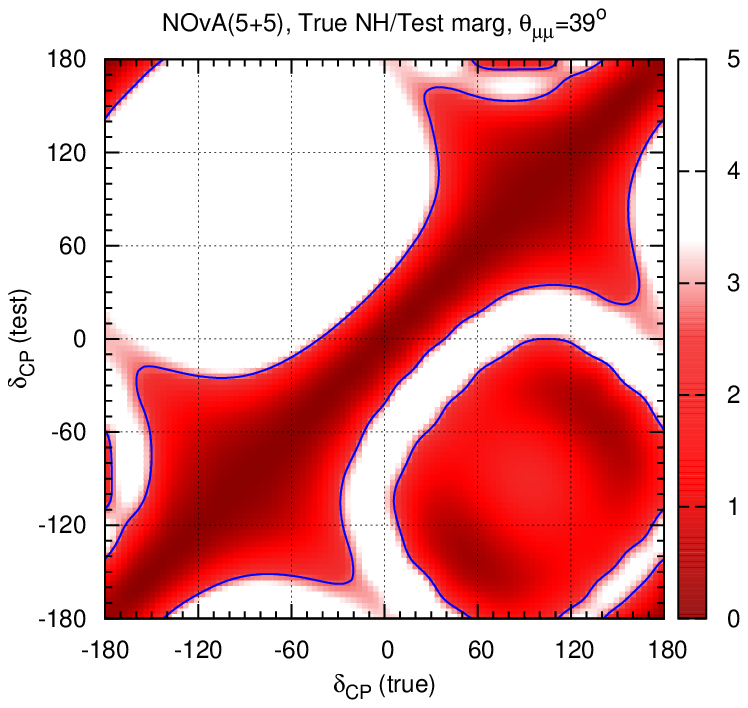, width=0.45\textwidth, bbllx=120, bblly=70, bburx=335, bbury=270,clip=}
\caption{{90$\%$ C.L. $\dcp$ precision for T2K (5+5) (left column)  and \nova
(5+5) (right column) 
for $\theta_{\mu\mu}=39^\circ$, $\sin^2 2\theta_{13} = 0.1$ and true NH. The three
panels in each column
correspond to test NH, test IH and marginalization over hierarchy.
}}
\label{t2k55nova55breakup}
\end{figure}

\begin{figure}[hbt]
\hspace{-0.2in}
\epsfig{file=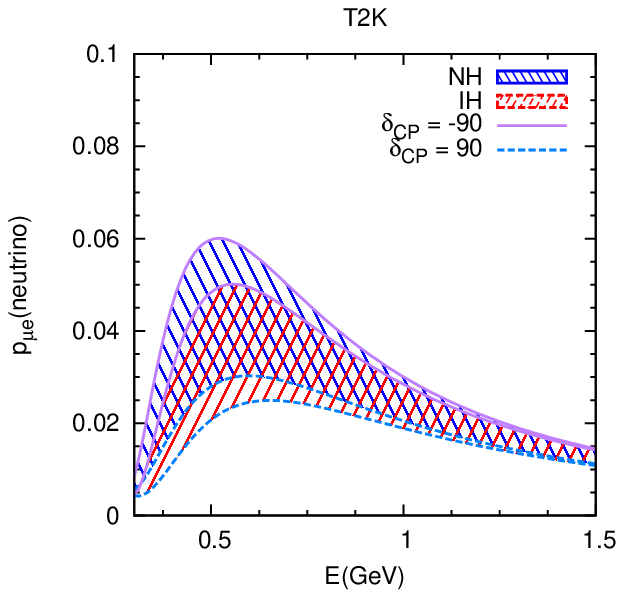, width=0.45\textwidth, bbllx=80, bblly=50, bburx=265, bbury=235,clip=}
\epsfig{file=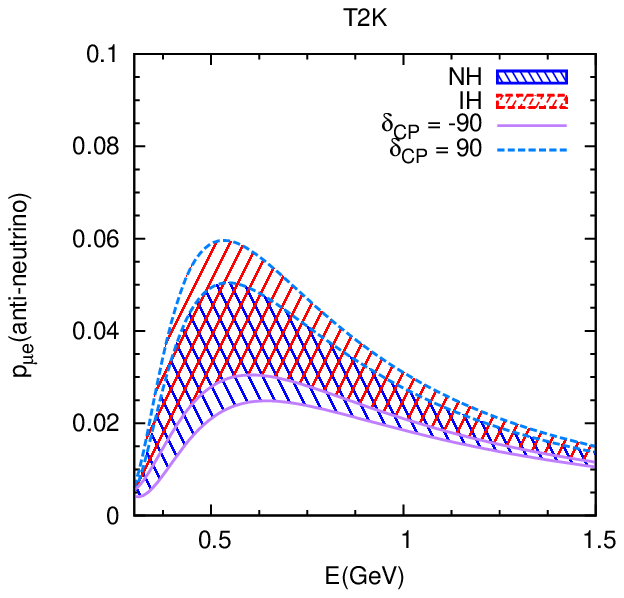, width=0.45\textwidth, bbllx=80, bblly=50, bburx=265, bbury=235,clip=} \\
\vspace{0.1in}
\epsfig{file=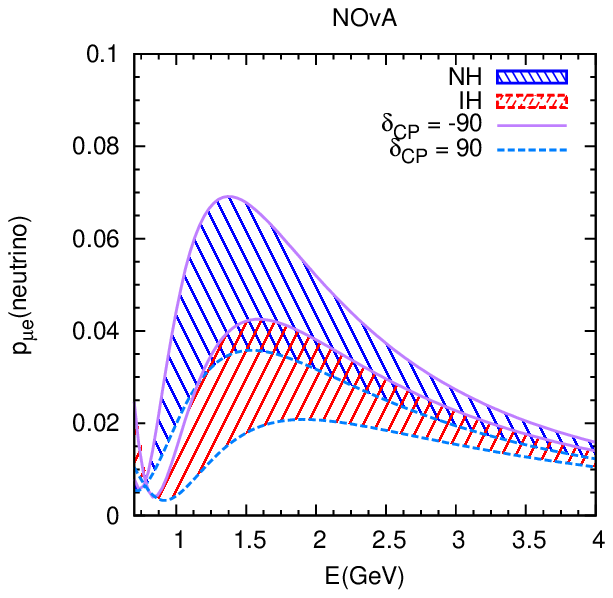, width=0.45\textwidth, bbllx=80, bblly=50, bburx=265, bbury=235,clip=}
\epsfig{file=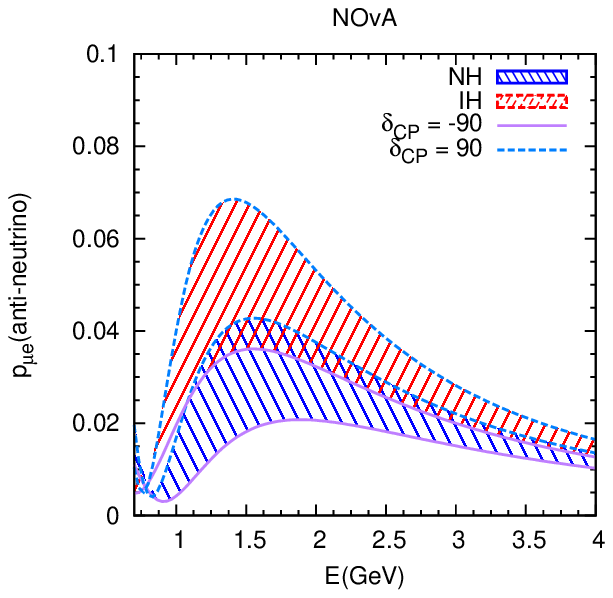, width=0.45\textwidth, bbllx=80, bblly=50, bburx=265, bbury=235,clip=}
\caption{{$\pmue$ energy spectrum for T2K (upper row) and \nova (lower row) for
neutrinos (left panel) and antineutrinos (right panel),
showing the bands for NH and IH when $\dcp$ is varied over the full range. The
curves for $\dcp=90^\circ$ and $-90^\circ$ are highlighted.
}}
\label{t2knovapmue}
\end{figure}
In this section we consider the possibility of a (5+5) run for T2K
in conjunction with \nova (5+5) run.  This is 
a possible extension beyond the projected timescale of the experiments.  
Fig.\ref{t2k55nova55}
illustrates the CP violation discovery potential, 90$\%$ C.L. $\dcp$ precision 
and 95$\%$ C.L. $\dcp$ precision for \nova (5+5) + T2K (5+5) 
for $\theta_{\mu\mu}=39^\circ$, $\sin^2 2\theta_{13} = 0.1$ and true NH (left panel)
or
true IH (right panel). It may be observed that in this case the discovery
potential rises to well above 3$\sigma$ for maximal CP violation in the
favourable half-plane, and stays above 3$\sigma$ between
$-120^\circ<{\rm{true}}~\dcp<-60^{\circ}$ for true NH and
$60^\circ<{\rm{true}}~\dcp<120^{\circ}$ for true IH. In the unfavourable
half-plane a 2$\sigma$ discovery signal is achieved over part of the true $\dcp$
range, but the discovery minima still occur with the wrong hierarchy. Similarly,
while the off-axis islands in the precision plot corresponding to the
wrong-hierarchy $\dcp$ solutions vanish at the level of 90$\%$ C.L., they are
still not ruled out at 95$\%$ C.L. { This shows the need for some additional 
input in order to resolve the hierarchy-$\dcp$ degeneracy in the
unfavourable half-plane.

It is worthwhile to analyze the relative contributions of \nova and T2K in this
case, where they have equal exposures with both neutrinos
and antineutrinos. While T2K has better statistics, \nova enjoys greater
hierarchy sensitivity due to a longer baseline and stronger earth matter
effects. To study this, we plot in Fig.\ref{t2k55nova55cpfraction} the allowed
fraction of $\dcp$ values at 90$\%$ C.L. for T2K (5+5) and \nova (5+5) as a function of true
$\dcp$. This quantity indicates the fraction of test $\dcp$ values which lie in
the allowed region for each specific value of true $\dcp$. Hence smaller values
of the allowed CP fraction signify better CP sensitivity. 

The figure is plotted for true NH. The three panels correspond to test NH, test
IH and a marginalization over hierarchy. It is observed that for a fixed NH,
\nova does slightly better than T2K. For test IH, \nova and T2K perform
similarly in the unfavourable half-plane (UHP), but \nova is much better than
T2K in the favourable half-plane (LHP) due to its superior hierarchy
sensitivity. However, with a marginalization over the unknown 
hierarchy, \nova does much
worse than T2K in the unfavourable half-plane. 

This anomalous feature can be explained from the 90$\%$ C.L. $\dcp$ precision
plots for T2K (5+5) and \nova (5+5)
(true NH) in Fig.\ref{t2k55nova55breakup}. The three panels in each column
correspond to test NH, test IH and marginalization over hierarchy.      
For fixed true and test NH (top row), T2K has a slightly larger allowed region
than \nova. For test IH (middle row), \nova does much better than 
T2K in terms of the allowed range covered. However, the allowed region of \nova
for test IH lies within the UHP of true $\dcp$ and LHP of test $\dcp$, which
is an excluded region for test NH.  Because of these disparate allowed regions,
with a marginalization over hierarchy (bottom row), \nova gives an 
allowed region along the axis as well as in the true UHP - test LHP region,
increasing its allowed CP fraction. On the other hand, for T2K, there are
significant overlaps between the allowed regions for test NH and test IH, and
the true UHP - test LHP range remains excluded in both cases. So a
marginalization over hierarchy does not cause as much of an increase in the
allowed CP fraction for T2K as it does for \nova. 

The reason for this difference in the behaviour of \nova and T2K can be seen at
the level of probabilities. Fig.\ref{t2knovapmue} depicts  
the $\pmue$ energy spectrum for the T2K and \nova baselines for neutrinos and
antineutrinos,
showing the bands for NH and IH when $\dcp$ is varied over the full range. The
curves for $\dcp=90^\circ$ and $-90^\circ$ are highlighted.
It is easy to see that due to the greater separation between the NH and IH bands
for \nova, the true NH - test IH case shows a clear degeneracy between the two
bands near true $\dcp=90^\circ$ and test $\dcp=-90^\circ$, leading to the true
UHP - test LHP allowed region in the \nova test IH precision figure. T2K has a
much greater overlap between the NH and IH bands, but in this case, the overlap
is more prominent in the regions of true UHP - test UHP and true LHP - test LHP,
corresponding to the allowed areas in these ranges in the T2K test IH precision
figure.
Hence in spite of the smaller allowed regions for \nova compared to T2K
especially for true NH/test IH, the location of the allowed regions
leads to an anti-synergistic combination for \nova (5+5), giving an overall poorer
CP sensitivity than T2K (5+5).

\section{CP violation discovery potential of T2K/\nova: synergies and dependence on parameters}

In this section, we study the behaviour of the CP violation discovery potential
as a function of
the neutrino parameters $\theta_{13}$, the neutrino mass
hierarchy and the octant of $\theta_{23}$. 
We also examine the synergy between the individual channels.
The discussion of synergies and parameter dependence here is
for the case T2K (5+0) + \nova (5+5), i.e. with a time frame till 2024.

\subsection{Synergy between appearance and disappearance channels of T2K/\nova:}

The event rates in T2K and \nova get contributions from both $P_{\mu \mu}$ 
and $P_{\mu e}$ channels.  
Due to the different behaviours of the two channels as a function of $\dcp$ and
other oscillation parameters,
there is a synergy between them which leads to an enhancement of the CP
violation discovery potential
of the combination. 
In Figure \ref{NovaT2Kappdisapp}, the CP violation discovery is plotted  
as a function of true $\dcp$ for the appearance and disappearance channels of
\nova and T2K for $\theta_{\mu\mu}^{tr}=39^\circ$, $\sin^2 2\theta_{13}^{tr} = 0.1$ and
true NH.
The following features can be observed:

\begin{enumerate}
\item The CP violation discovery potential principally arises from the
appearance channel of \nova/T2K, which is a function of $\pmue$,  owing to its
dependence on the quantity $\cos(\Delta+\dcp)$
in the sub-leading term of Eq.\ref{P-mue} as discussed in Section II.  
The disappearance channel offers a weaker $\dcp$ sensitivity through a
sub-leading dependence on $\cos \dcp$ \cite{akhmedov}.
The blue curve shows that by itself, the disappearance channel ($\pmumu$)
has negligible discovery potential.

\item Due to the different behaviours of the two channels as a function of
$\dcp$ and other oscillation parameters, there is a synergy between them which
leads to an enhancement of the CP violation discovery potential of the
combination. $\pmue$ is a function of both $\sin \dcp$ and $\cos \dcp$ while
$\pmumu$ depends only on $\cos \dcp$. From the green curve, it can be seen
that the discovery potential of the combination is significantly greater than
the sum of the discovery $\chi^2$ of the individual channels.

\item Both \nova and T2K experience this synergy between the appearance and
disappearance channels. In addition, there is a further enhancement of the
discovery potential when the two experiments are combined, as discussed in the
previous section.

\end{enumerate}

\begin{figure}[hbt]
\hspace{-0.2in}
\epsfig{file=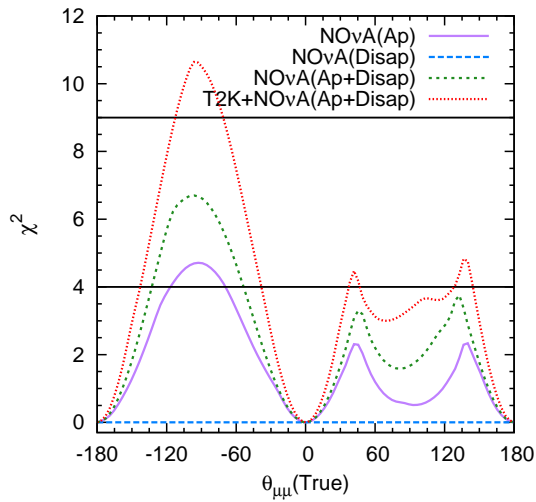, width=0.45\textwidth, bbllx=80, bblly=50, bburx=265, bbury=235,clip=}
\caption{{CP violation discovery as a function of true $\dcp$ for appearance(ap) and
disappearance(disap) channels of \nova (5+5) and T2K (5+0) for $\theta_{\mu\mu}^{tr}=39^\circ$, $\sin^2
2\theta_{13}^{tr} = 0.1$ and true NH.
}}
\label{NovaT2Kappdisapp}
\end{figure}

\subsection{Dependence on $\theta_{13}$:}

The behaviour of CP sensitivity as a function of $\theta_{13}$ can be understood by
looking at the $\theta_{13}$-dependence of the $\nu_\mu \to \nu_e$ oscillation probability $P_{\mu e}$.
As seen in Eqn.~\ref{P-mue}, $P_{\mu e}$ has a 
leading order term $\sim \sin^2 \theta_{13}$ that is independent of $\dcp$, and a sub-leading term 
$\sim \sin 2\theta_{13}$ that is a function of $\dcp$
 In calculating CP sensitivity $\chi^2$, the 
leading order $\dcp$-independent term cancels out from the true and test spectra in the numerator, 
but remains in the denominator. For illustrative purposes, the $\chi^2$ can be expressed as 
\begin{equation}
\chi^2 \sim \frac{P(\dcp) \sin^2 2\theta_{13}}{Q \sin^2 \theta_{13} + R(\dcp) \sin 2\theta_{13}} ~,
\end{equation} 
where $P$, $Q$, $R$ are functions of the other oscillation parameters 
apart from $\dcp$ and $\theta_{13}$. 
It is easy to show that for small values of $\theta_{13}$,
$\chi^2 \sim \theta_{13}$ which is an increasing function. 
It is also straightforward to consider the other limit,
 where $\theta_{13}$ is close to $90^\circ$. In 
this limit, $\chi^2 \sim  (90^\circ - \theta_{13})^2$ 
which decreases with $\theta_{13}$.
This feature can be 
understood qualitatively by noting that the leading order term is independent of 
$\dcp$ and therefore 
acts as a background to the CP signal \cite{lindner}. 
Therefore, CP sensitivity initially increases with $\theta_{13}$, 
peaks at an optimal value, and 
then decreases with $\theta_{13}$.
These features are reflected 
in Fig.\ref{theta13} where 
we plot the CP violation discovery potential of \nova+T2K
as a function of $\sin^2 2\theta_{13}^{tr}$ for two maximally CP-violating
values of true $\dcp$. We assume $\theta_{\mu\mu}^{tr} = 39^\circ$ and
a fixed normal
mass hierarchy. A marginalization over $\theta_{13}$ is done in the left panel
and $\theta_{13}$ is fixed to its true values in the right panel.
It can be seen that the discovery $\chi^2$ rises for very
small values of $\sin^2 2\theta_{13}$ and reaches its highest value in the
range $\sin^2 2\theta_{13} \sim 0.08 - 0.2$ before starting to drop off
gradually.  The vertical lines denote the current $\theta_{13}$ range ($\sin^2
2\theta_{13} = 0.07-0.13$).
This figure shows that the range of $\theta_{13}$ that nature has provided us
with is a fortuitous one, since it happens to lie 
in a region where the sensitivity to  CP violation is maximum 
with such experiments. 

Fig.\ref{th13} depicts the CP violation discovery as a function of true $\dcp$
for \nova+T2K
{{(true NH, $\theta_{13}$ and hierarchy marginalized, 
$\theta_{\mu\mu}^{tr}=39^\circ$) for two values of $\sin^2 2\theta_{13}^{tr}$ at the lower
and higher end of its present range and two values $\theta_{13}$ prior.}}  
It can be seen that in the favourable half-plane of $\dcp^{tr}$, there is a
slight increase in the $\chi^2$ with an increase in $\theta_{13}^{tr}$ in this
range, as can be predicted from Fig.\ref{theta13}. In the unfavourable
half-plane, there is again a complicated dependence of the discovery $\chi^2$ on
the intrinsic CP violation discovery of the experiments as well as their
hierarchy sensitivity, and since the latter increases significantly with
$\theta_{13}$, we observe a more definite improvement of the overall discovery
potential with increasing $\theta_{13}$. 

\begin{figure}[hbt]
\hspace{-0.2in}
\epsfig{file=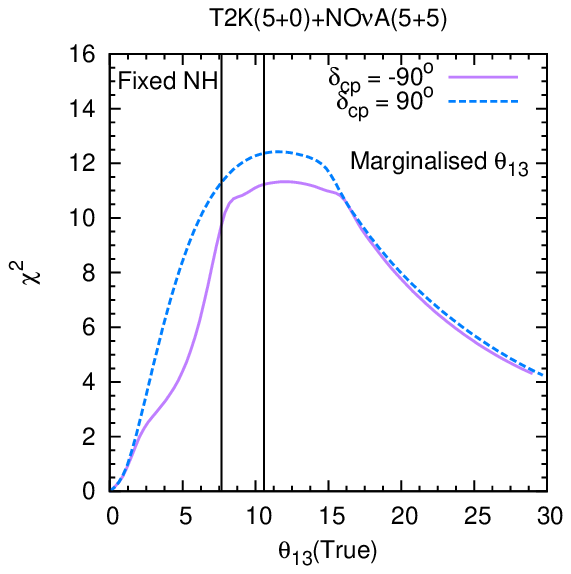, width=0.45\textwidth, bbllx=80, bblly=50, bburx=265, bbury=235,clip=}
\epsfig{file=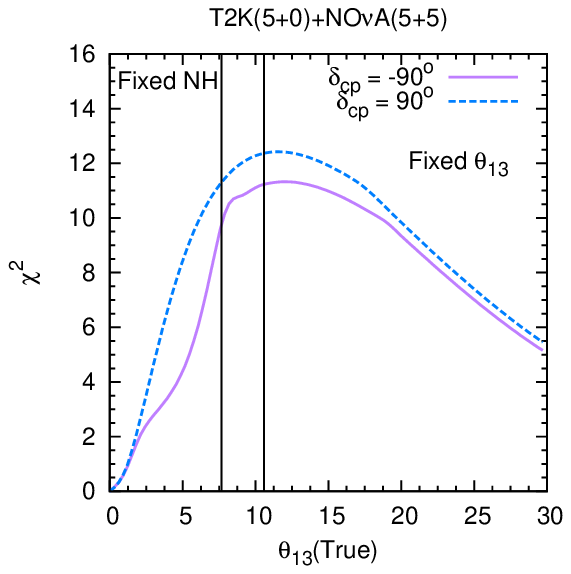, width=0.45\textwidth, bbllx=80, bblly=50, bburx=265, bbury=235,clip=}
\caption{{CP violation discovery potential of \nova+T2K as a function of true
$\theta_{13}$ for different values of true $\dcp$. $\theta_{\mu\mu}^{tr} = 39^\circ$
and a fixed NH is assumed. $\theta_{13}$ is marginalized in the left panel and
fixed in the right panel.
}}
\label{theta13}
\end{figure}

\begin{figure}[hbt]
\hspace{-0.2in}
\epsfig{file=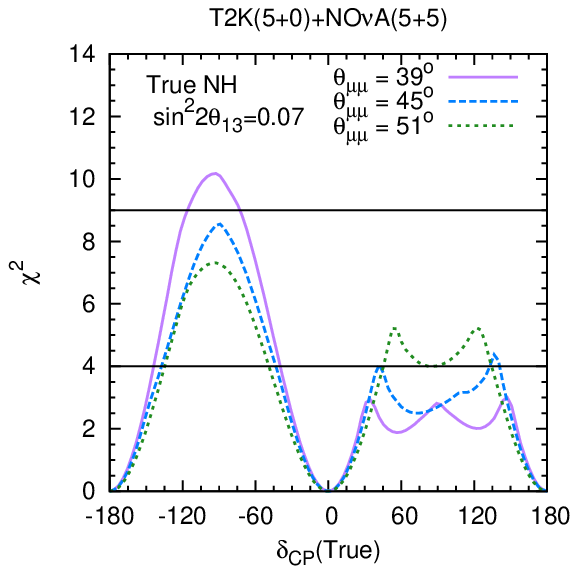, width=0.45\textwidth, bbllx=80, bblly=50, bburx=265, bbury=235,clip=}
\epsfig{file=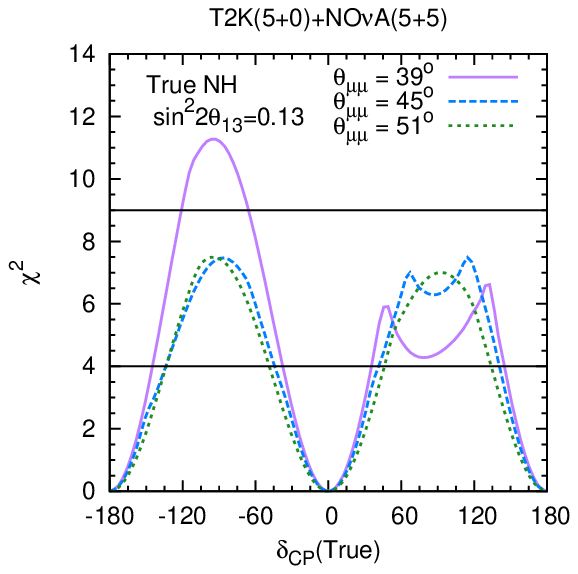, width=0.45\textwidth, bbllx=80, bblly=50, bburx=265, bbury=235,clip=}
\caption{{CP violation discovery as a function of true $\dcp$ for \nova+T2K
for three values of $\theta_{\mu\mu}^{tr}$, two values of $\sin^2 2\theta_{13}^{tr} =
0.07, 0.13$ and true NH, with a marginalization over the hierarchy and
$\theta_{13}$.
}}
\label{th13}
\end{figure}

\subsection{Dependence on the neutrino mass hierarchy:}

This aspect has been discussed in detail in \cite{cpv_ino}. 
Fig.\ref{hier} shows the CP violation discovery as a function of true $\dcp$ for
\nova+T2K
for three values of $\theta_{\mu\mu}^{tr}$, $\sin^2 2\theta_{13}^{tr} = 0.1$ and
true NH (left panel) or 
IH (right panel). As expected, there is a drop in the discovery $\chi^2$ in the
unfavourable half-plane 
in each case, i.e. in the UHP for true NH and in the LHP for true IH. In these
regions,
the discovery minima occur with the wrong hierarchy due to the hierarchy-$\dcp$
degeneracy,
and the discovery $\chi^2$ is a sum of the intrinsic discovery potential 
and the hierarchy sensitivity of \nova+T2K \cite{cpv_ino}. 

\begin{figure}[hbt]
\hspace{-0.2in}
\epsfig{file=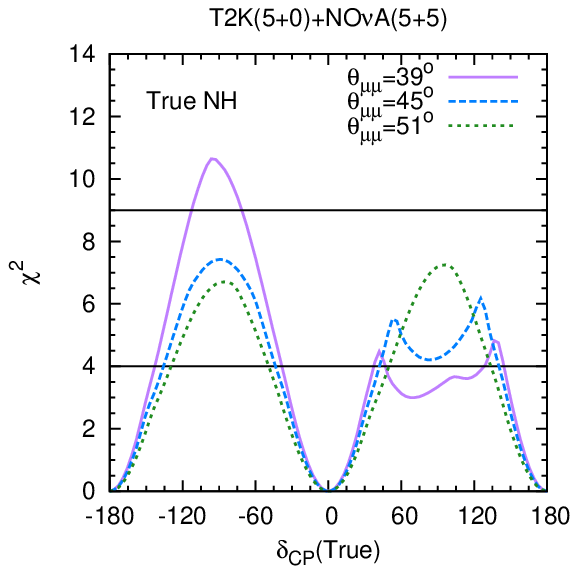, width=0.45\textwidth, bbllx=80, bblly=50, bburx=265, bbury=235,clip=}
\epsfig{file=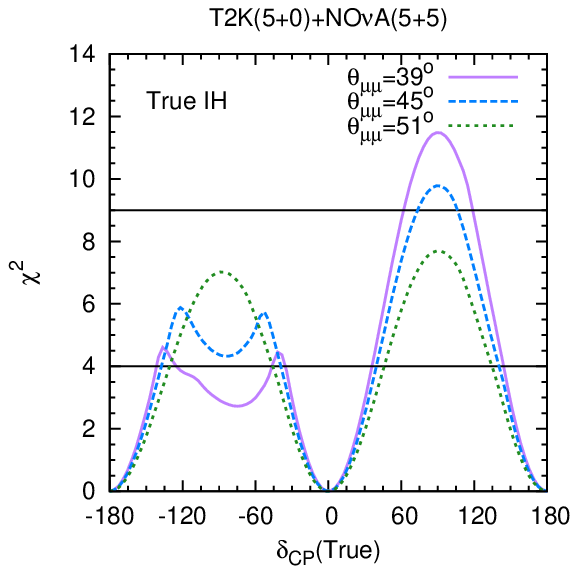, width=0.45\textwidth, bbllx=80, bblly=50, bburx=265, bbury=235,clip=}
\caption{{CP violation discovery as a function of true $\dcp$ for \nova+T2K
for three values
of $\theta_{\mu\mu}$, $\sin^2 2\theta_{13} = 0.1$ and true NH (left panel) or 
IH (right panel).
}}
\label{hier}
\end{figure}

\subsection{Dependence on $\theta_{\mu\mu}$ and octant:}

In Fig.\ref{hier}, we can also observe the dependence of the CP violation
discovery $\chi^2$ on the true value of $\theta_{\mu \mu}$.
When true $\dcp$ lies in the favourable half-plane, the discovery potential
decreases with increasing $\theta_{\mu \mu}$
in the current allowed range of $\theta_{\mu \mu}$.
In the unfavourable half-plane, the behaviour is more complicated since the
discovery minima lie in the wrong-hierarchy region for part of the range,
and the hierarchy sensitivity adds to the discovery $\chi^2$. The hierarchy
sensitivity is directly proportional to $\theta_{\mu\mu}$, and therefore
the overall CP violation discovery potential in these regions also increases
with $\theta_{\mu\mu}$.

As seen in Eqn.~\ref{P-mue}, $P_{\mu e}$ has a 
leading order  $\dcp$ independent term $\sim \sin^2 \theta_{\mu\mu}$ and a 
sub-leading $\dcp$ dependent term  
$\sim \sin 2\theta_{\mu\mu}$. This is similar to the   
$\theta_{13}$ behaviour.  Thus for smaller values of $\theta_{\mu\mu}$ 
the $\chi^2$ is expected to rise, reaching a peak  at an 
intermediate value of $\theta_{\mu\mu}$ and decreasing thereafter.
This is reflected in the left panel of
fig.\ref{theta23} where the CP violation discovery potential of \nova+T2K
is shown as a
function of true $\theta_{\mu\mu}$. This plot is drawn   
for two maximally CP-violating values of true
$\dcp$,  $\sin^2 2\theta_{13}^{tr} = 0.1$ and a fixed NH with 
test $\theta_{\mu \mu}$ fixed to its true value. 
The vertical lines give
the present 3$\sigma$ range of $\theta_{\mu\mu}$ ($\theta_{\mu\mu} =
35^\circ-55^\circ$).
Therefore,
as we increase $\theta_{\mu\mu}$ in its allowed range, we see a
drop of sensitivity.

The right panel of Fig. \ref{theta23} is obtained by marginalizing over the 
octant i.e assuming no prior knowledge of the octant in which $\theta_{\mu\mu}$ 
lies.  
We find that for $\theta_{\mu\mu}^{tr} < 40^\circ$ or $> 49^\circ$, there is no
effect of a marginalization over the octant. This is because the octant
sensitivity of
\nova + T2K is good enough (at least 2$\sigma$) in this range of
$\theta_{\mu\mu}^{tr}$ to rule out CP discovery solutions in the wrong octant
\cite{usoctant}.
The octant $\chi^2$ adds to the CP discovery $\chi^2$ in the wrong octant and
excludes any minima occurring in that region.
For $40^\circ < \theta_{\mu\mu}^{tr} < 49^\circ$, the octant sensitivity of
\nova+T2K is not high enough to exclude wrong-octant solutions,
and we see a wiggle in the discovery $\chi^2$ curves signaling the
octant-$\dcp$ degeneracy.
The behaviour is different for $\dcp^{tr} = \pm 90^\circ$,
since the LHP is favourable for resolving the octant-$\dcp$ degeneracy for true
HO and the UHP is favourable for true LO (in the neutrino mode, which gives the predominant contribution in these results).
This is illustrated in Fig.\ref{th23}, where
the discovery potential of \nova+T2K is plotted as a function of true $\dcp$ for
$\theta_{\mu\mu}^{tr} = 43^\circ$ (left panel) and $49^\circ$ (right panel) with and
without a marginalization over the octant.  $\sin^2 2\theta_{13}^{tr} = 0.1$ and
a fixed NH is assumed. These values of $\theta_{\mu\mu}^{tr}$ lie within the range
of unresolved octant-$\dcp$ degeneracy,
which shows up as a drop in the curve in the LHP for $\theta_{\mu\mu}^{tr} =
43^\circ$ and in the UHP for $\theta_{\mu\mu}^{tr} = 49^\circ$ when the octant is
assumed to be unknown, as expected from
 the above argument.  The favourable half-plane in each case suffers from no
degeneracy. We also see that the drop due to the octant degeneracy is greater in the case of 
$\theta_{\mu\mu}^{tr} = 43^\circ$ than for $49^\circ$ since the former value lies in the
central part of the degenerate region, while the latter is at the edge.

\begin{figure}[hbt]
\hspace{-0.2in}
\epsfig{file=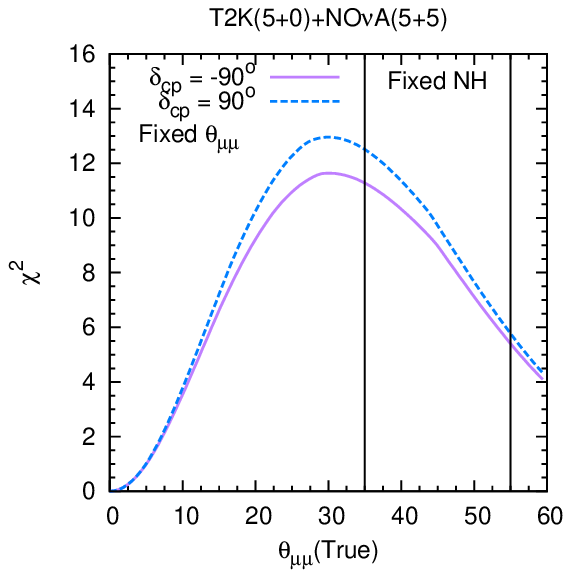, width=0.45\textwidth, bbllx=80, bblly=50, bburx=265, bbury=235,clip=}
\epsfig{file=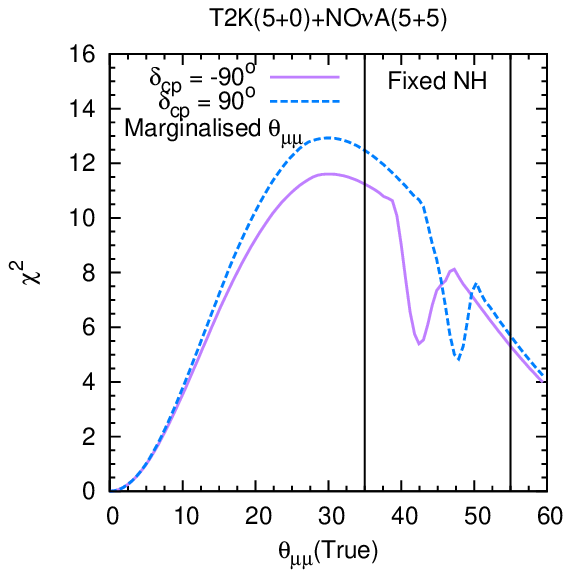, width=0.45\textwidth, bbllx=80, bblly=50, bburx=265, bbury=235,clip=}
\caption{{CP violation discovery potential of \nova+T2K as a function of true
$\theta_{\mu\mu}$ for two maximally CP-violating values of true $\dcp$. $\sin^2
2\theta_{13}^{tr} = 0.1$ and a fixed NH is assumed. The $\theta_{\mu\mu}$ octant is
fixed in the left panel and marginalized in the right panel.
}}
\label{theta23}
\end{figure}

\begin{figure}[hbt]
\hspace{-0.2in}
\epsfig{file=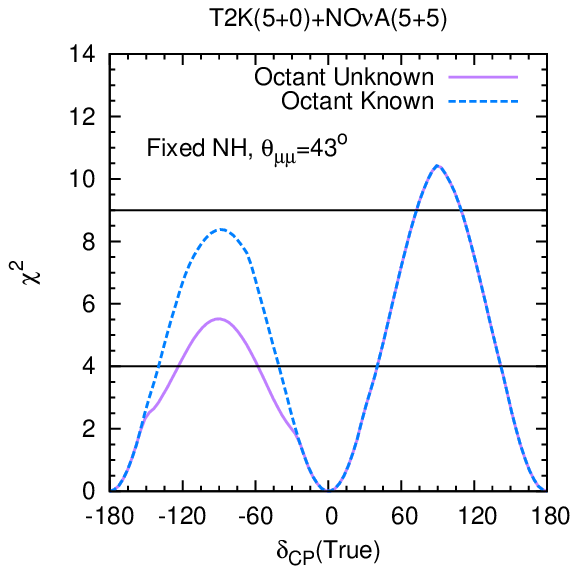, width=0.45\textwidth, bbllx=80, bblly=50, bburx=265, bbury=235,clip=}
\epsfig{file=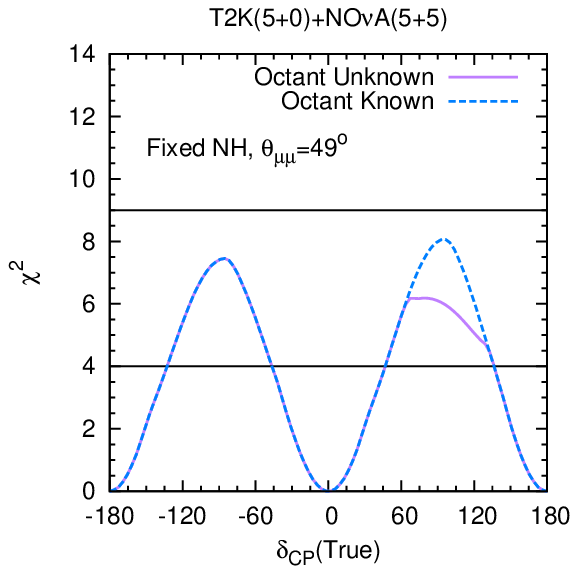, width=0.45\textwidth, bbllx=80, bblly=50, bburx=265, bbury=235,clip=}
\caption{{CP violation discovery potential of \nova+T2K as a function of true
$\dcp$ for $\theta_{\mu\mu}^{tr} = 43^\circ$ (left panel) and $49^\circ$ (right
panel), with and without a marginalization over the octant.  $\sin^2
2\theta_{13}^{tr} = 0.1$ and a fixed NH is assumed.
}}
\label{th23}
\end{figure}

\section{CP violation discovery and $\dcp$ precision with combined \nova, T2K
and atmospheric neutrinos}

{ In general, the CP sensitivity of 
atmospheric neutrino experiments are limited by the finite 
detector resolutions.  In particular, the angular resolutions 
need to be very good to have any intrinsic   
CP sensitivity from these experiments 
\cite{cpv_ino,cp_samanta,hagiwara_pingu}. 
It has been highlighted in \cite{cpv_ino}, that in spite of this limitation
the atmospheric neutrino experiments can play a crucial 
role in discovering CP violation in combination with 
the current generation LBL experiments T2K and \nova.  
The main reason for this is the ability of the atmospheric 
neutrino experiments to  
lift the hierarchy-$\dcp$ degeneracy
by excluding
discovery $\chi^2$ minima occurring with the wrong hierarchy in the unfavourable
half-plane of $\dcp^{tr}$}. 
This is achieved due to the
significant and largely $\dcp$-independent hierarchy sensitivity of atmospheric
neutrino experiments. 
This was demonstrated in 
in \cite{cpv_ino} taking ICAL@INO as the atmospheric neutrino detector. 
In this work we do a combined study for 
T2K + \nova + ICAL with different exposures and 
for both the CP violation discovery potential and the
$\dcp$ precision.

\subsection{CP sensitivity of ICAL}

In this sub-section we explore some details of   
the CP sensitivity of atmospheric neutrino
experiments. The main issue here is that the atmospheric neutrinos
come from all directions.  Hence 
these experiments face a further challenge of accurately reconstructing 
the direction apart from the energy.  We investigate how the intrinsic CP 
sensitivity of atmospheric neutrinos depend on the energy and angular 
resolutions and how much sensitivity can be achieved for an ideal detector.  

In Fig.\ref{res} the CP violation discovery potential of ICAL is plotted as a
function of the energy and angular resolution. 
The curve for angular(energy) resolution is plotted by varying the respective smearing widths  
between $3^\circ - 15^\circ$ ($3\% - 15\%$) while holding the 
energy(angular) resolution fixed at $10\%(10^\circ)$.
The figure illustrates the significant role played by the angular resolution of
an atmospheric neutrino detector in its CP sensitivity.
With present realistic values of detector smearing (15$\%$,15$^\circ$), the CP
sensitivity of such an experiment is  washed out 
by averaging over bins in energy and direction, due to the coupling between
$\dcp$ and $\Delta = \Delta_{31}L/4E$ 
in the term $\cos (\dcp + \Delta)$ in $\pmue$ \cite{cpv_ino}. With a hypothetical
improved angular resolution of $3^\circ$, the CP violation
discovery $\chi^2$ may reach values close to 1, going up to 5 for ideal detector
resolutions in both angle and energy.   

\begin{figure}[hbt]
\epsfig{file=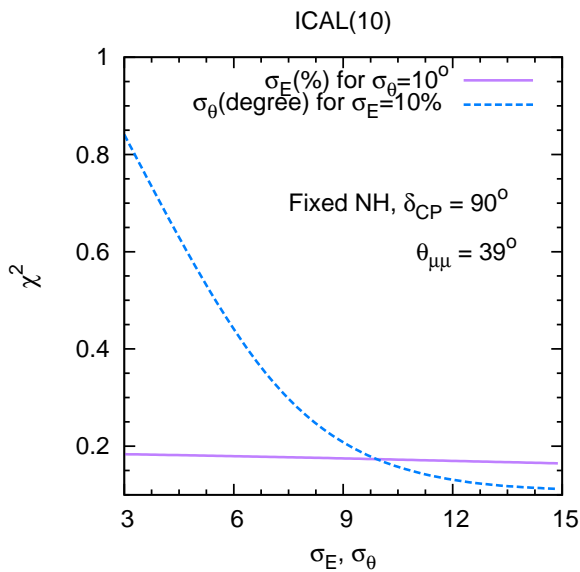, width=0.52\textwidth, bbllx=80, bblly=50, bburx=265, bbury=235,clip=}
\caption{{CP violation discovery potential of ICAL as a function of the detector
energy and angular resolutions for $\dcp^{tr} = 90^\circ$. $\theta_{\mu\mu}^{tr} =
39^\circ$ and a true NH is assumed. 
}}
\label{res}
\end{figure}

\subsection{CP sensitivity of T2K (5+0) with \nova (3+3) or (5+5) and ICAL 5 years (2024) or
10 years (2028)}

\begin{figure}
\hspace{-0.2in}
\epsfig{file=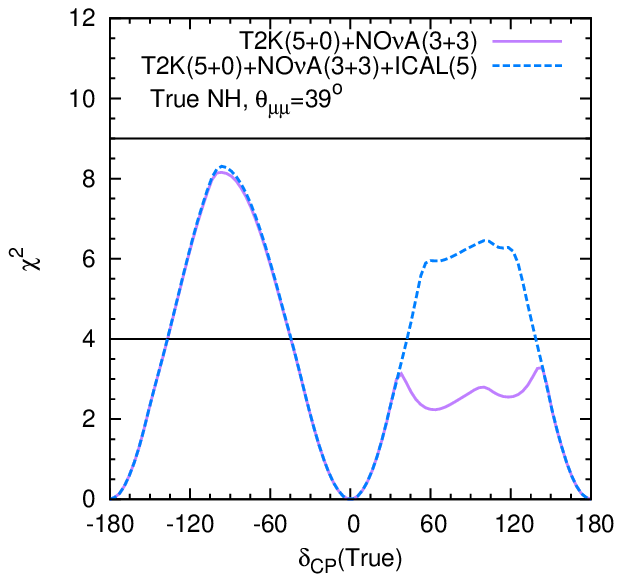, width=0.45\textwidth, bbllx=80, bblly=50, bburx=265, bbury=235,clip=}
\epsfig{file=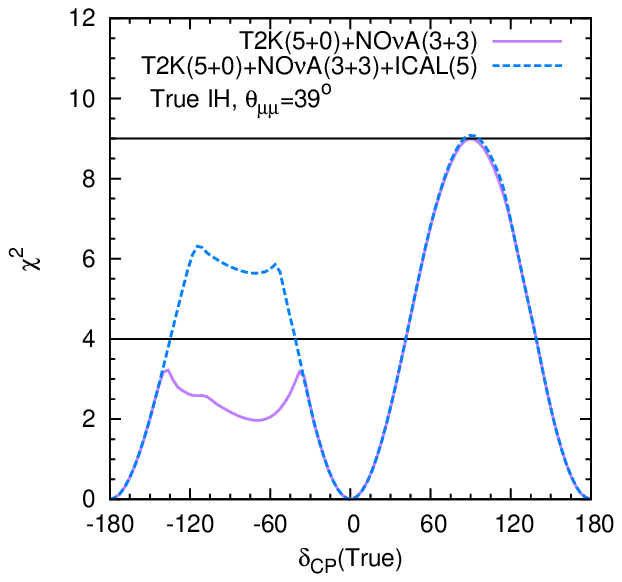, width=0.45\textwidth, bbllx=80, bblly=50, bburx=265, bbury=235,clip=} \\
\epsfig{file=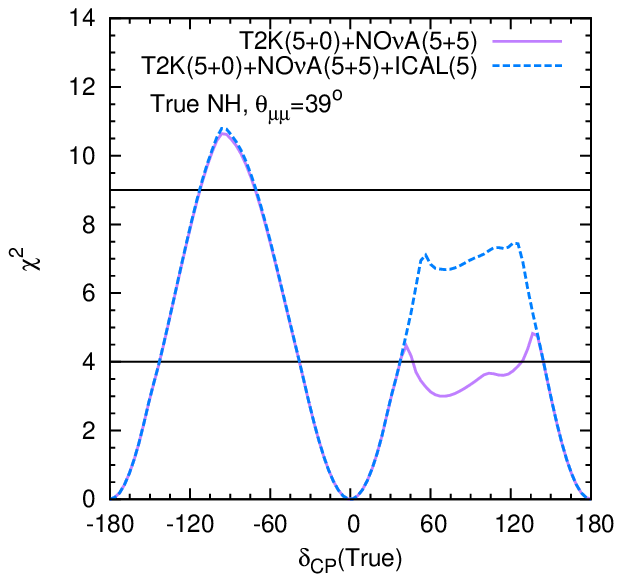, width=0.45\textwidth, bbllx=80, bblly=50, bburx=265, bbury=235,clip=}
\epsfig{file=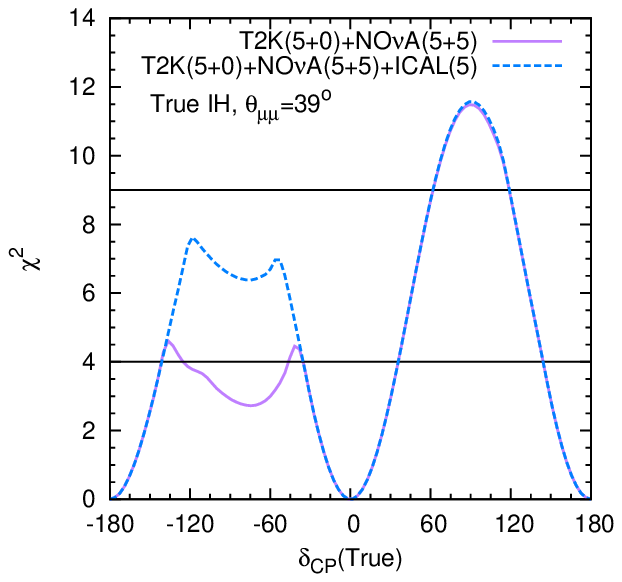, width=0.45\textwidth, bbllx=80, bblly=50, bburx=265, bbury=235,clip=} \\
\epsfig{file=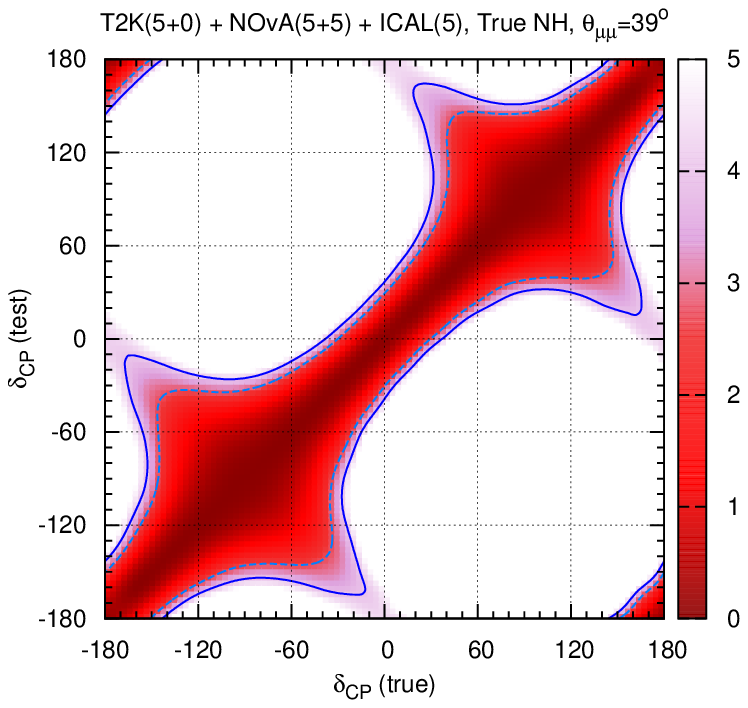, width=0.45\textwidth, bbllx=120, bblly=70, bburx=335, bbury=270,clip=}
\epsfig{file=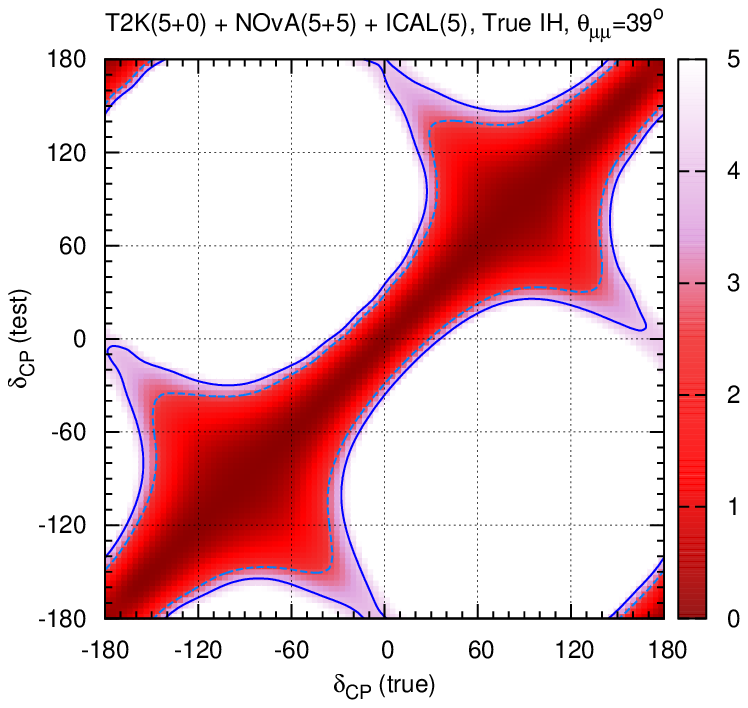, width=0.45\textwidth, bbllx=120, bblly=70, bburx=335, bbury=270,clip=}
\caption{{CP violation discovery (upper row) and 90 $\%$/95 $\%$ $\dcp$ precision (lower
row) for \nova (5+5) +T2K (5+0) + ICAL
(5 years)
for $\theta_{\mu\mu}=39^\circ$, $\sin^2 2\theta_{13} = 0.1$ and true NH (left panel)
or 
IH (right panel).
}}
\label{t2k50nova55ical5}
\end{figure}

{{Fig.\ref{t2k50nova55ical5} shows the CP violation discovery 
$\chi^2$ and  
90$\%$/95$\%$
C.L. $\dcp$ precision 
for \nova (5+5) + T2K (5+0) + ICAL (5 years). 
The discovery $\chi^2$ is also given for \nova (3+3).
 for $\theta_{\mu\mu}
= 39^\circ$, $\sin^2 2\theta_{13} = 0.1$, true NH (left panel)
and true IH (right panel). Fig.\ref{t2k50nova55ical10} plots the CP violation
discovery for \nova (5+5) + T2K (5+0) with ICAL (10 years)}}. 
The discovery plots show that with the addition of 5 years of ICAL data, while
the discovery $\chi^2$ in the favourable half-planes is unchanged compared to
Fig.\ref{t2k50nova55} as expected, there is a rise of about 3-4 in the $\chi^2$
values over a significant range in the unfavourable half-planes. The figure
tells us that the unfavourable half-plane still exhibits a hierarchy-$\dcp$
degeneracy and has discovery minima with the wrong hierarchy, but the
combination of the hierarchy sensitivity of ICAL (5 years) raises the discovery
potential to about 2.5$\sigma$ over the central part of the unfavourable
half-plane {{for the \nova (5+5) case}},
i.e. in the range $-120^\circ<{\rm{true}}~\dcp<-60^\circ$ (true IH) or
$60^\circ<{\rm{true}}~\dcp<120^\circ$ (true NH).
{{With \nova (3+3), the discovery potential reaches up to 2.5$\sigma$ for maximal CP violation
in both half-planes when ICAL 5 years data is added}}. 
From Fig.\ref{t2k50nova55ical10}, it can be seen that 10 years of ICAL data
provides a complete resolution of the hierarchy-$\dcp$ degeneracy and the
discovery potential goes up to 3$\sigma$ for maximal CP violation in both
favourable and unfavourable half-planes. 

\begin{figure}[hbt]
\hspace{-0.2in}
\epsfig{file=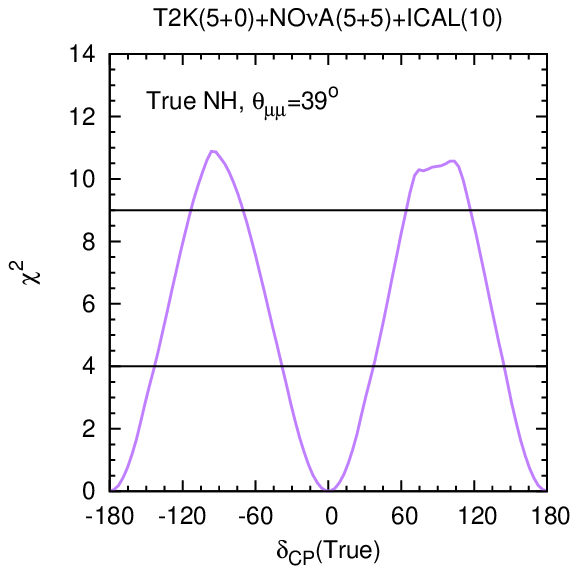, width=0.45\textwidth, bbllx=80, bblly=50, bburx=265, bbury=235,clip=}
\epsfig{file=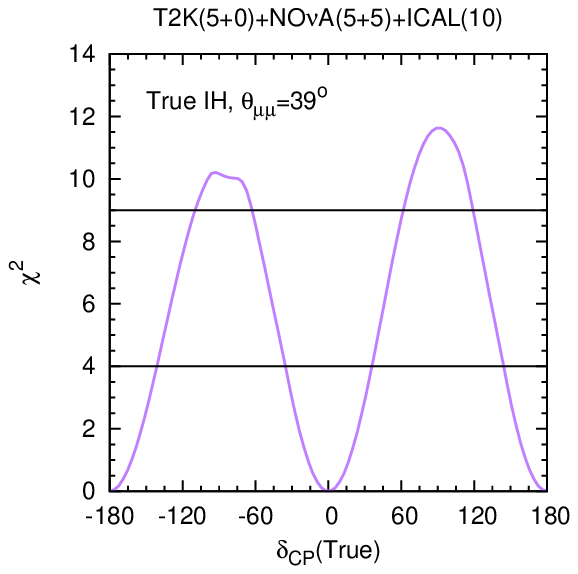, width=0.45\textwidth, bbllx=80, bblly=50, bburx=265, bbury=235,clip=} 
\caption{{CP violation discovery for \nova (5+5) +T2K (5+0) + ICAL (10 years)
for $\theta_{\mu\mu}=39^\circ$, $\sin^2 2\theta_{13} = 0.1$ and true NH (left panel)
or IH (right panel).
}}
\label{t2k50nova55ical10}
\end{figure}

Regarding the $\dcp$ precision, it may be recalled from Fig.\ref{t2k50nova55}
(lower row) that with \nova (5+5) + T2K (5+0) alone, the 90$\%$/95$\%$ C.L. allowed
regions include some islands in the off-axis region,
i.e. with $\dcp^{tr}$ in the UHP and $\dcp^{test}$ in the LHP for true NH and
vice versa for true IH. 
These correspond to the CP minima occurring with the wrong hierarchy due to the
hierarchy-$\dcp$ degeneracy.   
From the precision plots in Fig.\ref{t2k50nova55ical5}, it can be observed that
these wrong-hierarchy solutions go away at both 90$\%$ and 95$\%$ C.L. when
atmospheric neutrino information from ICAL (5 years) is combined, since
this degeneracy is resolved by the addition of hierarchy sensitivity from ICAL. 
Thus the combination of atmospheric neutrino experiments with \nova/T2K can aid
the potential
for $\dcp$ measurement of the long-baseline experiments by curtailing the 
allowed range, and for this purpose ICAL data of 5 years is enough to exclude
the wrong-hierarchy solutions up to 95$\%$ C.L. or about 2$\sigma$.

\begin{figure}[hbt]
\epsfig{file=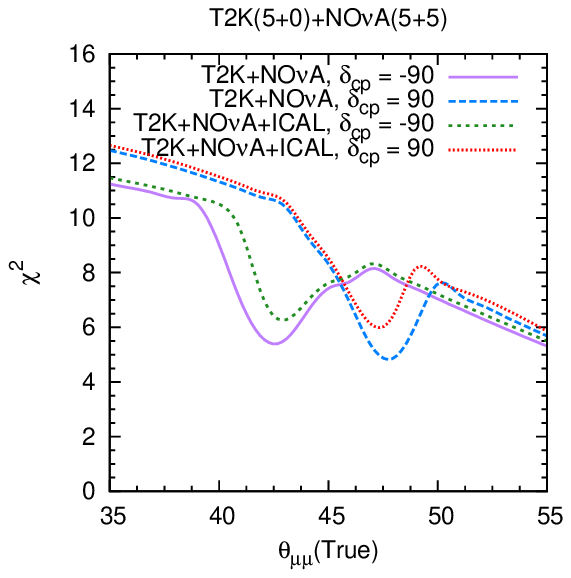, width=0.52\textwidth, bbllx=80, bblly=50, bburx=265, bbury=235,clip=}
\hspace*{0.5ex}
\caption{{CP violation discovery potential of \nova (5+5) +T2K (5+0) + ICAL (10
years) as a function of true $\theta_{\mu\mu}$ for two maximally CP-violating values
of true $\dcp$. $\sin^2 2\theta_{13}^{tr} = 0.1$ and a fixed NH is assumed and
$\theta_{\mu\mu}$ is marginalized over both octants.
}}
\label{theta23ical}
\end{figure}

We also study what happens to the octant-$\dcp$ degeneracy when ICAL is combined
with \nova and T2K. Fig.\ref{theta23ical} shows the CP violation discovery
potential of \nova (5+5) + T2K (5+0) with and without ICAL (10 years) as a function
of true $\theta_{\mu\mu}$ for true $\dcp = \pm 90^\circ$ for $\sin^2 2\theta_{13}^{tr}
= 39^\circ$ and a fixed NH, with a marginalization over $\theta_{\mu\mu}$. Comparing
with Fig.\ref{theta23},   
we see that the wiggle in the $40^\circ < \theta_{\mu\mu}^{tr} < 49^\circ$ range
corresponding to the octant-$\dcp$ degeneracy is reduced in amplitude and
restricted to the range $41^\circ < \theta_{\mu\mu}^{tr} < 48^\circ$ when ICAL data
is added. In Fig.\ref{th23ical}, the
 discovery potential of \nova + T2K  with and without ICAL (10 years) is plotted
as a function of true $\dcp$ for $\theta_{\mu\mu}^{tr} = 43^\circ$ (left panel) and
$49^\circ$ (right panel) with and without a marginalization over the octant, for
$\sin^2 2\theta_{13}^{tr} = 0.1$ and a fixed NH. These values of
$\theta_{\mu\mu}^{tr}$ lie within the range of unresolved octant-$\dcp$ degeneracy,
even with the combination of ICAL, but an improvement in the discovery $\chi^2$
is seen in the unfavourable half-plane in each case when ICAL is
added. Since the drop due to the octant degeneracy is greater for
$\theta_{\mu\mu}^{tr} = 43^\circ$ than for $49^\circ$, the addition of 
ICAL data entirely overcomes
the degeneracy and compensates for the drop in the $49^\circ$ case, while 
for $43^\circ$ there is only a partial improvement in the $\chi^2$ even when ICAL
data is added. 

\begin{figure}[hbt]
\hspace{-0.2in}
\epsfig{file=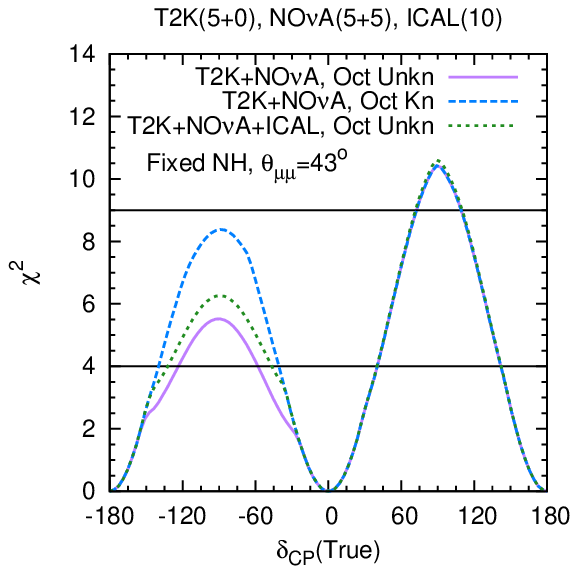, width=0.45\textwidth, bbllx=80, bblly=50, bburx=265, bbury=235,clip=}
\epsfig{file=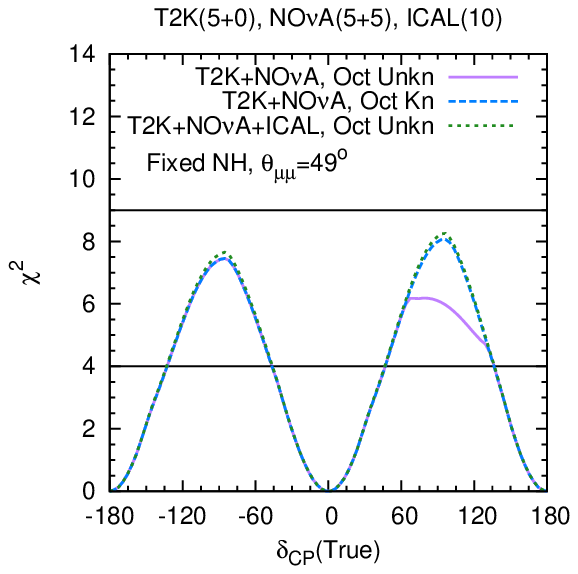, width=0.45\textwidth, bbllx=80, bblly=50, bburx=265, bbury=235,clip=} 
\caption{{CP violation discovery potential of \nova (5+5) + T2K (5+0) + ICAL (10
years) as a function of true $\dcp$ for $\theta_{\mu\mu}^{tr} = 43^\circ$ (left
panel) and $49^\circ$ (right panel), with and without a marginalization over the
octant.  $\sin^2 2\theta_{13}^{tr} = 0.1$ and a fixed NH is assumed.
}}
\label{th23ical}
\end{figure}

The effect of ICAL information on the octant-$\dcp$ degeneracy is more modest
than that for the hierarchy-$\dcp$ degeneracy since the octant sensitivity of
ICAL is not as good as its hierarchy sensitivity \cite{usoctant}. It is still
helpful to an extent since, like the hierarchy sensitivity, the octant
sensitivity is nearly independent of $\dcp$.  


\subsection{CP sensitivity of T2K (5+5) with \nova (5+5) and ICAL 5 years (2024) or
10 years (2028)}

Finally, we examine the benefits of adding ICAL to the {{projected}} combination of
T2K (5+5) + \nova (5+5). Fig.\ref{t2k55nova55} showed that while this combination
provides good discovery potential ($> 3\sigma$) over the central part of the
favourable half-plane, the unfavourable half-plane still suffers from the
hierarchy-$\dcp$ degeneracy and barely reaches a discovery potential of
2$\sigma$ over its central region. Further, the wrong-hierarchy solutions in the
$\dcp$ precision figure get ruled out at 90$\%$ C.L. but not at 95$\%$ C.L.

\begin{figure}[hbt]
\hspace{-0.2in}
\epsfig{file=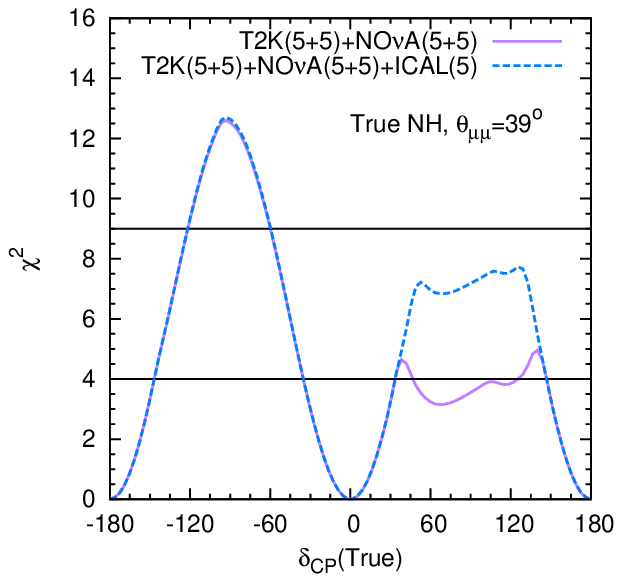, width=0.45\textwidth, bbllx=80, bblly=50, bburx=265, bbury=235,clip=}
\epsfig{file=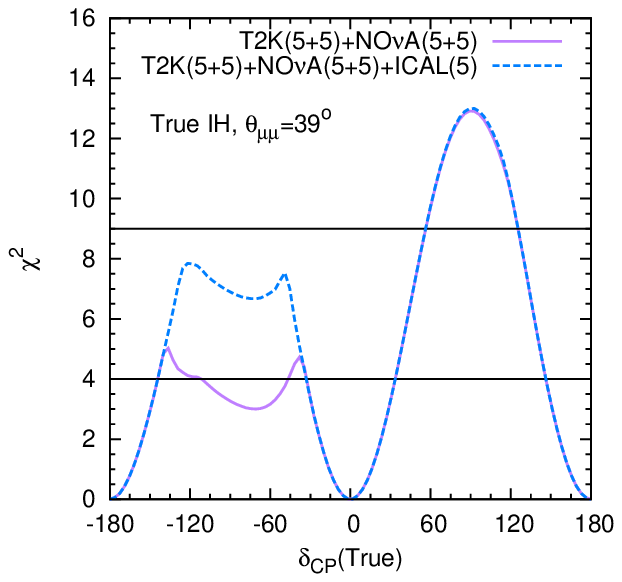, width=0.45\textwidth, bbllx=80, bblly=50, bburx=265, bbury=235,clip=} \\
\vspace{0.2in}
\epsfig{file=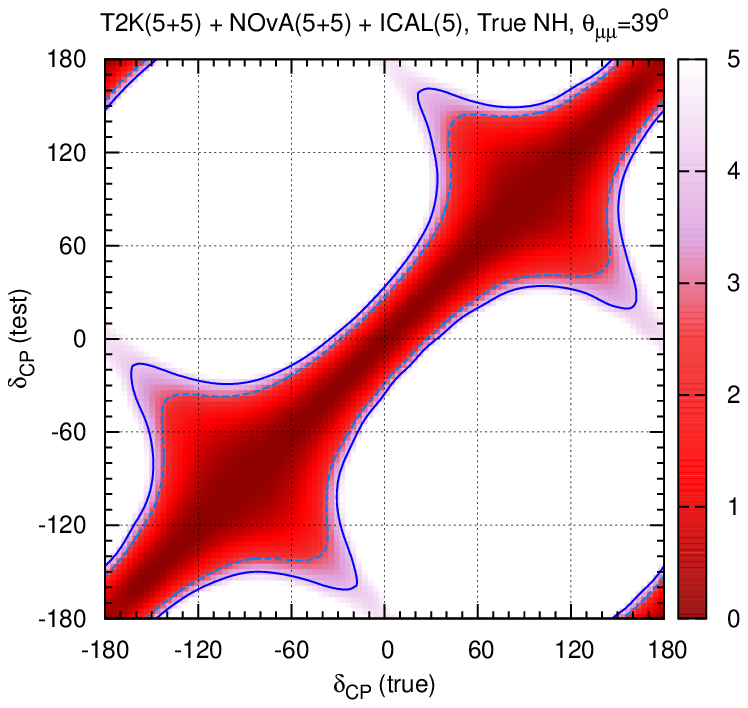, width=0.45\textwidth, bbllx=120, bblly=70, bburx=335, bbury=270,clip=}
\epsfig{file=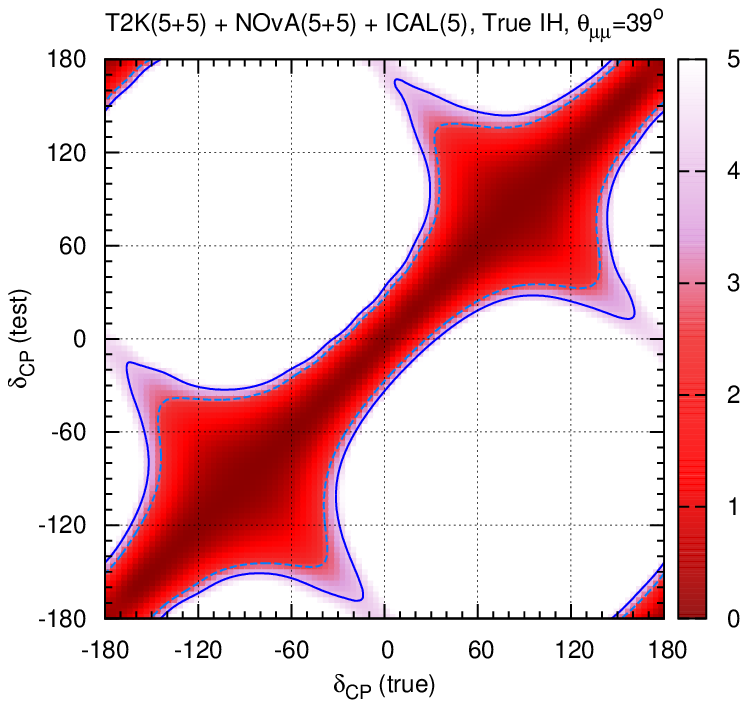, width=0.45\textwidth, bbllx=120, bblly=70, bburx=335, bbury=270,clip=}
\caption{{CP violation discovery (upper row) and 90$\%$/95$\%$ C.L. $\dcp$ precision
(lower row) for \nova (5+5) +T2K (5+5) + ICAL (5 years)
for $\theta_{\mu\mu}=39^\circ$, $\sin^2 2\theta_{13} = 0.1$ and true NH (left panel)
or 
IH (right panel).
}}
\label{t2k55nova55ical5}
\end{figure}

\begin{figure}[hbt]
\hspace{-0.2in}
\epsfig{file=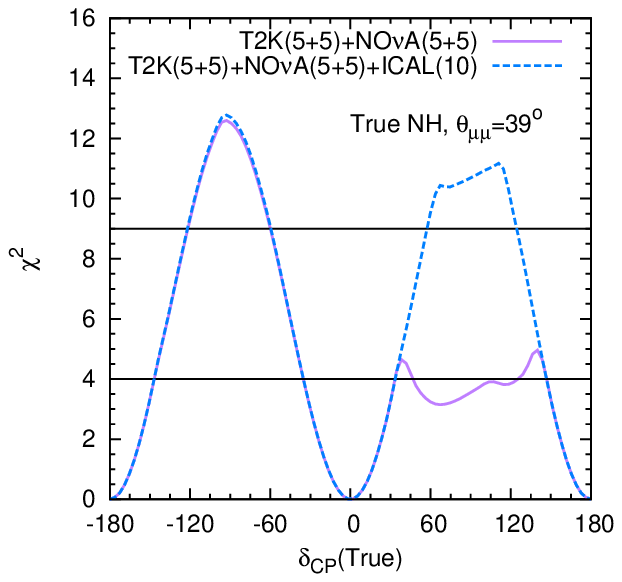, width=0.45\textwidth, bbllx=80, bblly=50, bburx=265, bbury=235,clip=}
\epsfig{file=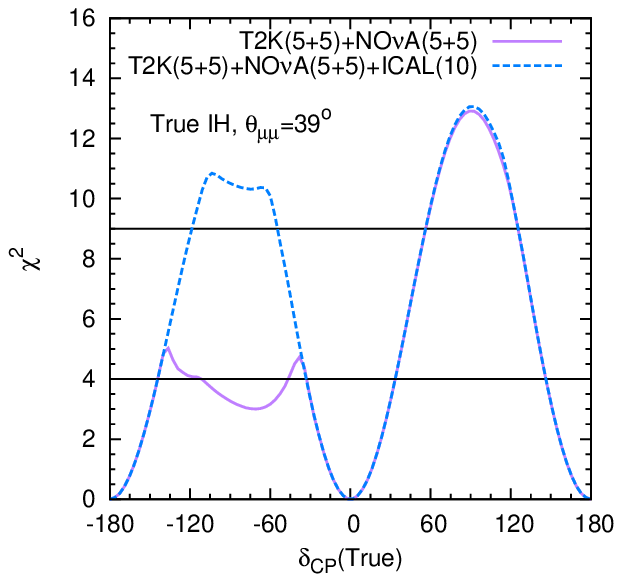, width=0.45\textwidth, bbllx=80, bblly=50, bburx=265, bbury=235,clip=} 
\caption{{CP violation discovery for \nova (5+5) +T2K (5+5) + ICAL (10 years)
for $\theta_{\mu\mu}=39^\circ$, $\sin^2 2\theta_{13} = 0.1$ and true NH (left panel)
or 
IH (right panel).
}}
\label{t2k55nova55ical10}
\end{figure}

In Fig.\ref{t2k55nova55ical5} we plot the CP violation discovery and 95$\%$ C.L.
$\dcp$ precision for \nova (5+5) +T2K (5+5) + ICAL (5 years). 
Fig.\ref{t2k55nova55ical10} depicts the discovery potential for ICAL (10 years).
The figures illustrate that with the addition of 5 years of ICAL data, the
discovery potential in the unfavourable half-plane is improved to about
2.7$\sigma$ over the range $-130^\circ<{\rm{true}}~\dcp<-50^\circ$ (true IH) or
$50^\circ<{\rm{true}}~\dcp<130^\circ$ (true NH), even though the
hierarchy-$\dcp$ degeneracy is still present. The favourable half-plane, as
expected, is unaffected by the addition of ICAL. Also, the small off-axis
allowed regions at 95$\%$ C.L. in the precision plot for T2K (5+5) + \nova (5+5) get
excluded when ICAL (5 years) is added. Hence the combination of ICAL constrains
$\dcp$ with a higher level of sensitivity. With 10 years of ICAL data, the
hierarchy-$\dcp$ degeneracy is fully resolved and the discovery potential of the
\nova+T2K+ICAL combination achieves values above 3$\sigma$ 
over the ranges $-120^\circ<{\rm{true}}~\dcp<-60^\circ$ as well as 
$60^\circ<{\rm{true}}~\dcp<120^\circ$, i.e. in both the favourable and
unfavourable half-planes for both hierarchies.
Thus the addition of ICAL can provide a more consistent signature of CP
violation and a more constrained measurement of $\dcp$.

\section{Conclusions}

Measuring CP violation in the lepton sector is one of the most challenging problems 
today. We have performed a systematic chronological study of the CP sensitivity of the
current and upcoming long-baseline experiments T2K and \nova and the atmospheric
neutrino experiment with a prototype of ICAL@INO. We analyze the synergies
between these set-ups which may aid in CP violation discovery and a precision
measurement of $\dcp$. This has been done for different combinations of these
experiments which will be achievable at progressive points in time in the near
future. The main role of the atmospheric data is to rule out the 
wrong hierarchy solutions which increases the CP sensitivity in the 
unfavourable parameter regions for T2K/\nova.  
Usually the analysis of CP sensitivity is done assuming the hierarchy/octant 
to be known
--in which case the wrong hierarchy/wrong octant solutions
are excluded a priori. 
We show how a realistic atmospheric neutrino 
experiment can achieve this and quantify the  
exposure which enables one to disfavour the wrong hierarchy  and/or wrong octant
solutions. 
Below we list the salient features of our results. 

\vskip.2cm

{\bf{Study of synergies and parameter dependence:}}

\vskip.2cm

\begin{itemize}

\item While the CP sensitivity principally arises from the appearance channel of
\nova/T2K, the appearance and disappearance channels are synergistic due to
their different dependences on $\dcp$.  $\pmue$ depends on $\dcp$ through
the quantity $\cos(\Delta+\dcp)$, while $\pmumu$ only has a $\cos \dcp$ 
dependence.
Thus their combination gives a CP sensitivity significantly higher than the sum
of sensitivities of the two channels. 

\item 
The results for a combination of T2K and \nova display  hierarchy-$\dcp$ 
degeneracy. 
This is manifested 
as a drop in the CP violation discovery potential in the unfavourable
half-plane 
of $\dcp$, i.e. the UHP ($0-180^\circ$)for true NH and the LHP ($-180^\circ-0$) for 
true IH. 

\item There is also a 
degeneracy of $\dcp$ with the octant. However because of significant 
octant sensitivity of the T2K + \nova combination, this occurs over a restricted range of $\theta_{\mu\mu}$ around the maximal value. 
For example, for 
 a T2K (5+0) + \nova (5+5) combination, the degeneracy with the
octant occurs over the range $40^\circ < \theta_{\mu\mu}^{tr} < 49^\circ$
The degeneracy shows up as a drop in the discovery potential in LHP for true 
LO ($\theta_{\mu\mu}<45^\circ$) and in the UHP for true HO ($\theta_{\mu\mu}>45^\circ$).

\item 
Although a non-zero $\theta_{13}$ is essential for any measurement of 
$\dcp$, large values of this parameter can also impede the CP sensitivity 
\cite{minakata}. 
This is because of the presence of the 
$\dcp$ independent  leading term in $\pmue \sim \sin^2\theta_{13}$, 
which can act as a background for the sub-dominant $\dcp$ dependent term.  
However we note that for smaller values of $\theta_{13}$ the CP-discovery 
$\chi^2 \sim \theta_{13}$ and hence increases with $\theta_{13}$. 
On the other hand, for larger values of $\theta_{13}$ the CP-discovery $\chi^2
\sim (90^\circ - \theta_{13})^2$ which 
decreases with $\theta_{13}$. The discovery $\chi^2$ 
attains its highest value in the range $\sin^2 2\theta_{13}
\sim 0.08 - 0.2$. This tells us that the range of
$\theta_{13}$ provided by nature lies in an optimal region which is favourable
for CP sensitivity with such experiments. 

\end{itemize}

\vskip.2cm

{\bf{Chronological study:}} 

\vskip.2cm

{{In Table \ref{tablesummary} we summarize the maximum values of CP violation discovery potential in the unfavourable half-plane of true $\dcp$, and the percentage of true $\dcp$ values capable of giving a CP violation discovery signal at 2$\sigma$ and 3$\sigma$, for different combinations of the experiments T2K, \nova and ICAL at progressive points of time over the next 15 years. The following observations can be made from these results:}}

\begin{itemize}

\item By 2016, T2K is expected to have an effective 5-year run with 10$^{21}$
pot/year. We consider the cases of a (5+0) versus a (3+2) run, and find that with
T2K alone, a (3+2) run provides a better CP sensitivity than a neutrino only (5+0)
run, due to the complementary behaviour of the neutrino and antineutrino
probabilities which partially resolves the hierarchy-$\dcp$ degeneracy in the
favourable half-plane of $\dcp$. 

\item By 2020, \nova will complete a (3+3) run. We combine this with the T2K
results for (3+2) and (5+0) and find that the combination offers similar CP
sensitivity in both cases. This is because \nova, with its combined
neutrino-antineutrino run, helps in resolving the hierarchy-$\dcp$ degeneracy in
the favourable half-plane and overrides the necessity of resolving it with T2K.
Thus a neutrino-only T2K run proves to be as efficient towards CP sensitivity as
a combined (3+2) run when it is taken in tandem with \nova. In this way the
combination of T2K and \nova provides a synergy, apart from the improved
sensitivity of the combination purely due to the increased statistics and
exposure. 

\item By 2024, \nova may have a (5+5) run. Combining this with T2K (5+0) adds to the
CP sensitivity due to the higher \nova exposure, and can provide a CP violation
discovery potential of up to 3$\sigma$ in the favourable half-plane and up to
2$\sigma$ at some points in the unfavourable half-plane. The $\dcp$ precision
determination is also improved but still displays some additional allowed
regions in $\dcp$ corresponding to the wrong-hierarchy solutions.

\item We also look at an extended (5+5) run of T2K, and consider it with \nova
(5+5). In this case the CP
violation discovery potential rises well above 3$\sigma$ for maximal CP
violation in the favourable half-plane. The unfavourable half-plane gives a
discovery signal of 2$\sigma$ over parts of the true $\dcp$ range, but the
discovery minima still occur with the wrong hierarchy. In the $\dcp$ precision
plots, the wrong-hierarchy allowed regions are ruled out at 90$\%$ C.L. but not
at 95$\%$ C.L. 

\begin{table}
\begin{center}
\begin{tabular}{|c || c | c | c | c |} \hline
      Experiment (timeline) & 
 \multicolumn{2}{c}{\parbox{0.2\textwidth}{max $\chi^2$ in \\ UVHP \qquad FVHP \qquad }} 
& \multicolumn{2}{c}{\parbox{0.2\textwidth}{$\dcp$ fraction for CPV \\$2\sigma$ \qquad $3\sigma$ \qquad }}  \\
        \hline
 T2K (3+2) (2016) & \parbox{0.1\textwidth}{$0.9$} &\parbox{0.1\textwidth}{$3.3$}& \parbox{0.1\textwidth}{$-$} & $-$   \\ \hline
 T2K (5+0) (2016) & $1.2$ & 0.8 & $-$ & $-$    \\ \hline 
 T2K (3+2) + \nova (3+3) (2020) & $3.1$ & 7.5 & $24\%$ & $-$    \\ \hline
 T2K (5+0) + \nova (3+3) (2020)  & $3.3$ & 8.2 & $25\%$ & $-$   \\ \hline
 T2K (5+0) + \nova (5+5) (2024)  & $4.8$ & 10.7 & $36\%$ & $11\%$  \\ \hline
 T2K (5+5) + \nova (5+5) (2024) & $4.9$ &12.5 & $41\%$ & $17\%$  \\ \hline
 T2K (5+0) + \nova (3+3) + ICAL 5 (2024) & $6.4$ & 8.3 & $52\%$ & $-$  \\ \hline
 T2K (5+0) + \nova (5+5) + ICAL 5 (2024) & $7.4$ & 10.8 & $60\%$ & $12\%$  \\ \hline
 T2K (5+5) + \nova (5+5) + ICAL 5 (2024) & $7.7$ & 12.7 & $62\%$ & $17\%$  \\ \hline
 T2K (5+0) + \nova (5+5) + ICAL 10 (2028) & $10.7$ & 11.0 & $60\%$ & $27\%$  \\ \hline
 T2K (5+5) + \nova (5+5) + ICAL 10 (2028) & $11.1$ & 12.7 & $62\%$ & $36\%$  \\ \hline  
          \hline
\end{tabular}
\caption{\small Values of maximal CP violation discovery $\chi^2$ in the favourable and unfavourable half-planes (FVHP and UVHP) and percentage of true $\dcp$ values allowing CP violation discovery at 2$\sigma$/3$\sigma$
for combinations of experiments at different chronological points. Here $\theta_{\mu\mu}^{tr} = 39^\circ$, $\sin^2 2\theta_{13} = 0.1$ and true NH.}
\label{tablesummary}
\end{center}
\end{table}

\item Finally we look at a combination of ICAL@INO with \nova and T2K, and find that
it resolves many of the issues with degeneracy observed in the \nova+T2K
results. By 2024, ICAL will have at least 5 years of data. With a T2K (5+0) +
\nova (5+5) + ICAL (5 years) combination, the CP violation discovery potential
still exhibits a hierarchy-$\dcp$ degeneracy and has discovery minima with the
wrong hierarchy, but due to the hierarchy sensitivity of ICAL, the discovery
potential is raised to about 2.5$\sigma$ over the central region of the
unfavourable half-plane. The favourable half-plane is unaffected by the addition
of ICAL. If we consider the situation in 2028, when ICAL is expected to have 10
years of data, the  T2K (5+0) + \nova (5+5) + ICAL (10 years) combination completely
resolves this degeneracy and the discovery potential goes up to 3$\sigma$ in
both the favourable and unfavourable half-planes. The wrong-hierarchy allowed
regions in the precision plots that are present for  T2K (5+0) + \nova (5+5) at
90$\%$ C.L. go 
away at both 90$\%$ and 95$\%$ C.L. with the T2K (5+0) + \nova (5+5) + ICAL (5
years) combination.  

\item Combining \nova and T2K with ICAL (10 years) also gives a modest
improvement in lifting the octant-$\dcp$ degeneracy, reducing the range of its effect
and improving the discovery potential in the unfavourable half-plane. The
advantage in this case is less than for the hierarchy-$\dcp$ degeneracy since
the octant sensitivity of ICAL is not as good as its hierarchy sensitivity.

\item For T2K (5+5) + \nova (5+5), a combination with ICAL (5 years) improves the
discovery potential to about 2.7$\sigma$ over the central part of the
unfavourable half-plane but the hierarchy-$\dcp$ degeneracy is still present.
Also, the small off-axis allowed regions at 95$\%$ C.L. in the precision plot
for T2K (5+5) + \nova (5+5) get excluded when ICAL (5 years) is added. With 10 years
of ICAL data, the hierarchy-$\dcp$ degeneracy is fully resolved and the
discovery potential of the \nova+T2K+ICAL combination achieves values above
3$\sigma$ over the central part of both the favourable and unfavourable
half-planes. Thus the addition of ICAL can provide a more consistent signature
of CP violation and a more constrained measurement of $\dcp$.

\end{itemize} 

{{In conclusion, the combination of T2K and \nova can provide reasonable CP 
sensitivity for some values of neutrino parameters but is severely compromised 
in this regard in other ranges. The addition of atmospheric neutrino information 
bearing uniform hierarchy sensitivity may be crucial in measuring $\dcp$ and 
detecting CP violation in case nature has chosen the parameter values unfavourable 
for LBL experiments. This fact has valuable ramifications for current experiments 
as well as for designing future LBL experiments like LBNO \cite{lbno}, where the 
inclusion of atmospheric neutrino data can significantly influence the exposures 
required for giving a high CP sensitivity over all allowed parameter values}}. 

\section*{Acknowledgements}

We thank Sanjib Kumar Agarwalla and Suprabh Prakash for useful discussions.

\bibliography{cpv2013.bib}

\end{document}